\documentclass[journal]{IEEEtran}
\usepackage{amssymb}
\usepackage[cmex10]{amsmath}
\usepackage{stfloats}
\usepackage{graphicx}
\usepackage{subfigure}
\usepackage{tabularx}
\usepackage{epsfig,epsf,color,balance,cite}
\usepackage{verbatim}
\usepackage{url}
\usepackage{bm}
\newtheorem{theorem}{Theorem}

\newtheorem{lemma}{Lemma}
\usepackage{algorithm}
\usepackage{algorithmic}
\begin{document}

% paper title
\title{Robust Beamforming with Pilot Reuse Scheduling in a Heterogeneous Cloud Radio Access Network}

\author{
\IEEEauthorblockN{Hao Xu\IEEEauthorrefmark{0},
                  Cunhua Pan\IEEEauthorrefmark{0},
                  Wei Xu, \emph{Senior Member, IEEE}\IEEEauthorrefmark{0},
                  Gordon L. St{\"u}ber, \emph{Fellow, IEEE}\IEEEauthorrefmark{0},
                  Jianfeng Shi\IEEEauthorrefmark{0}
 and
                  Ming Chen\IEEEauthorrefmark{0}
                  }
 \thanks{Copyright (c) 2015 IEEE. Personal use of this material is permitted. However, permission to use this material for any other purposes must be obtained from the IEEE by sending a request to pubs-permissions@ieee.org. This work was in part supported by the NSFC (Nos. 61372106, 61471114, \& 61221002), NSTMP under 2016ZX03001016-003, the Six Talent Peaks project in Jiangsu Province under GDZB-005, Science and Technology Project of Guangdong Province under Grant 2014B010119001, the Scholarship from the China Scholarship Council (No. 201606090039), Program Sponsored for Scientific Innovation Research of College Graduate in Jiangsu Province under Grant KYLX16\_0221, and the Scientific Research Foundation of Graduate School of Southeast University under Grant YBJJ1651. \emph{(Corresponding author: Cunhua Pan.)}}

 \thanks{H. Xu, W. Xu, J. Shi and M. Chen are with the National Mobile Communications Research
Laboratory, Southeast University, Nanjing 210096, China (e-mail: xuhao2013@seu.edu.cn; wxu@seu.edu.cn; shijianfeng@seu.edu.cn; chenming@seu.edu.cn).

 C. Pan is with School of Electronic Engineering and Computer Science, Queen Mary University of London, London E1 4NS, U.K. (e-mail: c.pan@qmul.ac.uk).

 G. L. St{\"u}ber is with the Faculty of Electrical and Computer Engineering, Georgia Institute of Technology, Atlanta, GA 30332 USA (e-mail: stuber@ece.gatech.edu).

}}

% make the title area
\maketitle

\begin{abstract}
%\boldmath
This paper considers a downlink ultra-dense heterogeneous cloud radio access network (H-CRAN) which guarantees seamless coverage and can provide high date rates. In order to reduce channel state information (CSI) feedback overhead, incomplete inter-cluster CSI is considered, i.e., each remote radio head (RRH) or macro base station (MBS) only measures the CSI from user equipments (UEs) in its serving cluster. To reduce pilot consumption, pilot reuse among UEs is assumed, resulting in imperfect intra-cluster CSI. A two-stage optimization problem is then formulated. In the first stage, a pilot scheduling algorithm is proposed to minimize the sum mean square error (MSE) of all channel estimates. Specifically, the minimum number of required pilots along with a feasible pilot allocation solution are first determined by applying the Dsatur algorithm, and adjustments based on the defined level of pilot contamination are then carried out for further improvement. Based on the pilot allocation result obtained in the first stage, the second stage aims to maximize the sum spectral efficiency (SE) of the network by optimizing the beam-vectors. Due to incomplete inter-cluster CSI and imperfect intra-cluster CSI, an explicit expression of each UE's achievable rate is unavailable. Hence, a lower bound on the achievable rate is derived based on Jensen's inequality, and an alternative robust transmission design (RTD) algorithm along with its distributed realization are then proposed to maximize the derived tight lower bound. Simulation results show that compared with existing algorithms, the system performance can be greatly improved by the proposed algorithms in terms of both sum MSE and sum SE.

\end{abstract}

\IEEEpeerreviewmaketitle

\section{Introduction}
\label{section1}
According to the forecast in \cite{index2016global}, there will be 1.5 mobile devices per capita and the monthly global mobile data traffic will surpass 49 exabytes by 2021. To meet the continuously growing demand for ubiquitous high-speed wireless access, a 1000 times capacity boost is thus expected in the fifth-generation (5G) network compared to the current fourth-generation network \cite{andrews2014will}. To realize this 5G vision, cloud radio access network (C-RAN) has been recognized as a promising solution \cite{wu2015cloud}. In the C-RAN architecture, a baseband unit (BBU) pool with powerful computation capability acts as a cloud data center, and remote radio heads (RRHs) configured only with some radio-frequency functionalities are connected to the BBU pool through optical fiber fronthaul links. Due to the simplified functionalities, RRHs can be densely and distributedly deployed to improve network access in conventional cellular networks, especially in hot spots with a large number of user equipments (UEs) like hospitals, shopping malls, etc. In addition, different from untra-dense small cell networks that suffer from cochannel interference \cite{chen2016user}, interference mitigation can be effectively realized in a C-RAN by applying Coordinated Multi-Point Transmission/Reception (CoMP) thanks to the powerful BBU pool \cite{dai2014sparse}.

Since a C-RAN is mainly adopted to provide high data rates in hot spots, real-time voice service and control signalling are not efficiently supported. If UEs move fast, the switching speed of the RRH service in a C-RAN is relatively high, resulting in a high signalling exchange load. On the other hand, the fronthaul capacity of a C-RAN is usually limited, making it difficult to serve all UEs in a network. Hence, lack of high power nodes may make it hard to ensure backward compatibility with the existing cellular systems \cite{Dahrouj2015Resource, Gerasimenko2015Cooperative}. In order to deliver the overall control signalling and guarantee seamless coverage, an advanced architecture, known as a heterogeneous cloud radio access network (H-CRAN), was proposed to combine the advantages of both C-RAN and heterogeneous networks \cite{peng2014heterogeneous, peng2015system, Peng2016Energy}. In an H-CRAN, besides the BBU pool and RRHs, macro base stations (MBSs) are also included. The BBU pool and MBSs are interfaced via backhaul links for coodination. Hence, the delivery of control and broadcast signalling can be shifted from RRHs to MBSs to alleviate the capacity and time delay on the fronthaul \cite{li2016energy, peng2015contract}. Unnecessary handover and re-association can then be avoided.

In H-CRANs, cochannel interference suppression is an important technical issue. In C-RANs, only interference among RRHs exists and it can be effectively mitigated via centralized processing at the BBU pool. While in H-CRANs, when RRHs and MBSs operate on the same time-frequency resource block (RB), a UE served by RRHs (a MBS) will not only experience intra-tier interference from RRHs (MBSs), but also suffer additional inter-tier interference from MBSs (RRHs). Hence, the signal-to-interference-plus-noise ratio (SINR) expressions of UEs contain more cochannel interference terms, making the problem more challenging. In \cite{li2016energy, peng2015contract, peng2015energy}, interference suppression problems in H-CRANs were studied aiming at different design metrics. Specifically, reference \cite{li2016energy} aimed to maximize the average throughput and maintain the network stability by traffic admission control, user association and resource allocation. In \cite{peng2015contract}, a contract-based interference coordination framework was proposed to mitigate the inter-tier interference between RRHs and MBSs. Reference \cite{peng2015energy} aimed to maximize the energy efficiency (EE) of an H-CRAN by resource assignment and power control. All these works, however, focused on single-input single-output (SISO) networks. By considering multi-antenna MBSs and multi-antenna RRHs, the performance of H-CRANs can be further enhanced by applying beamforming techniques.

Downlink beamforming design has been widely studied in both the conventional cellular networks \cite{christensen2008weighted, shi2011iteratively, wang2013energy} and C-RANs \cite{shi2014group, dai2016energy, pan2017joint}. Perfect global channel state information (CSI) was assumed available for system performance analysis and optimization. However, it is in practice difficult to obtain perfect CSI of all links due to channel estimation and quantization errors. In addition, feeding back the CSI from all UEs to each RRH or MBS requires excessive overhead, which may  easily overwhelm the capacity of the wireless radio interface, especially in ultra-dense C-RANs \cite{caire2010rethinking, lakshmana2016precoder}. Hence, transmission design based on incomplete CSI has drawn great attention recently \cite{Zhang2017Sum, Cirik2016Transceiver, Filippou2013A, Filippou2015Coordinated, Shen2012Distributed, shi2014csi, pan2017jointjsac}. In \cite{Zhang2017Sum} and \cite{Cirik2016Transceiver}, a multi-user massive multiple-input multiple-output (MIMO) network and a full-duplex MIMO cognitive radio system are respectively considered, and beamforming vectors were designed under channel uncertainties. In \cite{Filippou2013A, Filippou2015Coordinated, Shen2012Distributed}, distributed transmit beamforming was studied with imperfect CSI under different scenarios. Specifically, references \cite{Filippou2013A} and \cite{Filippou2015Coordinated} considered a cognitive radio network, while \cite{Shen2012Distributed} considered a multicell cellular system. In \cite{shi2014csi} and \cite{pan2017jointjsac}, incomplete CSI was assumed in C-RANs, i.e., each RRH only estimated the CSI from UEs in its serving cluster (named intra-cluster CSI). As for the UEs outside the serving cluster, it was assumed that the RRH only had the large-scale channel gains (named inter-cluster CSI). However, perfect intra-cluster CSI was still assumed in \cite{shi2014csi} and \cite{pan2017jointjsac}. To this end, orthogonal training has to be adopted for channel estimation. In this case, the length of pilot overhead increases linearly with the number of UEs, which could be unaffordable for ultra-dense networks.

One promising way to reduce pilot overhead is allowing pilots to be reused among UEs. Pilot reuse design has been extensively studied in massive MIMO networks \cite{yin2013coordinated, ashikhmin2012pilot, zhu2015graph, you2015pilot, yin2016robust} as well as device-to-device (D2D) underlaid systems \cite{liu2015pilot, liu2016pilot, xu2017pilot}. In particular, references \cite{yin2013coordinated, ashikhmin2012pilot, zhu2015graph} considered pilot reuse in multiple-cell scenarios, i.e., UEs in the same cell use orthogonal pilots, and the same set of pilots are reused in different cells. In \cite{you2015pilot} and \cite{yin2016robust}, it was shown that due to the uncorrelation feature of massive MIMO antennas, the same pilot could be reused by UEs with different angular positions. In \cite{liu2015pilot, liu2016pilot, xu2017pilot}, pilots were allowed to be reused by D2D pairs. Since RRHs are usually located dispersively and they use low power for short-distance transmission, the pilot reuse by UEs far away from each other would cause marginal pilot contamination. This paper mainly focuses on pilot scheduling and robust beamforming design for an ultra-dense H-CRAN. To the best of the authors' knowledge, this problem has not yet been studied. The main contributions of this paper are summarized as follows:

\hangafter=1
\setlength{\hangindent}{1.9em}
$\bullet$ This paper considers an ultra-dense H-CRAN, where RRHs are mainly used to provide high data rates and MBSs are deployed for guaranteeing seamless coverage as well as control signalling delivery. Different from the SISO scenarios in \cite{li2016energy, peng2015contract, peng2015energy}, it is assumed that both RRHs and MBSs are equipped with multiple antennas. In order to reduce CSI feedback overhead, incomplete CSI is considered, i.e., each RRH as well as MBS only measures the CSI from UEs in its serving cluster while tracks the large-scale channel gains of UEs outside its serving cluster. Let RUE and BUE represent the UEs served by RRHs and the MBS, respectively. To reduce pilot overhead, pilots are allowed to be reused by RUEs in different RRH clusters. Based on these settings, a two-stage optimization, i.e., pilot scheduling and robust transmission design, is considered to enhance the network performance.

\hangafter=1
\setlength{\hangindent}{1.9em}
$\bullet$ In the first stage, a problem aiming to minimize the sum mean square error (MSE) of all channel estimates is formulated. To distinguish the channels from different RUEs, it is assumed that the RUEs served by the same RRH apply orthogonal pilots for channel estimation. Upon this constraint, a minimum number of pilots applies. By constructing an undirected graph to describe this constraint and employing the Dsatur algorithm in \cite{brelaz1979new}, which aims to color the vertices of an undirected graph with the minimum number of different colors, the minimum number of pilots can be obtained. Since this algorithm only takes into account the constraint that RUEs served by the same RRH apply orthogonal training, while ignores the objective function, i.e., minimizing the sum MSE of channel estimation, it outputs a feasible pilot allocation solution, which may not be satisfactory. Hence, it is necessary to adjust the pilots allocated to each RUE by the Dsatur algorithm. A pilot scheduling algorithm (PSA) is thus proposed to further mitigate pilot contamination resulting from pilot reuse.

\hangafter=1
\setlength{\hangindent}{1.9em}
$\bullet$ The second stage aims to maximize the sum spectral efficiency (SE) of the network by optimizing the beam-vectors under incomplete CSI. Since each RRH or MBS has only imperfect intra-cluster CSI, it is difficult to obtain explicit expressions of the achievable rates of RUEs and BUEs. Lower bounds on the achievable rates are derived using the Jensen's inequality. Then, instead of directly solving the original problem, the achievable rate of each UE in the objective function is replaced with the lower bound. It is shown that the data rate lower bound of either an RUE or a BUE can be regarded as the rate of a mobile user in an equivalent downlink multiple-input single-output (MISO) interfering network. An alternative robust transmission design (RTD) algorithm along with its distributed realization are then provided to obtain a suboptimal solution.

\hangafter=1
\setlength{\hangindent}{1.9em}
$\bullet$ In the simulation part, the performances of the proposed algorithms are illustrated and compared in terms of both sum MSE and sum SE. Simulation results show that the sum MSE of channel estimation can be effectively suppressed by the proposed PSA. Compared with the existing schemes which assume perfect CSI, the sum SE of the network can be significantly increased by pilot reuse and the proposed RTD algorithm.

The rest of this paper is organized as follows. In Section~II, the signal transmission of an ultra-dense H-CRAN and the estimation of intra-cluster channels are presented. In Section~III, a pilot scheduling algorithm is proposed to minimize the sum MSE of channel estimation. In Section IV, an alternative algorithm along with its distributed realization are provided to maximize the sum SE of the network by optimizing the transmit beam-vectors. Numerical results are presented in Section V before conclusions in Section VI.

This paper follows commonly used notations. $\mathbb R$ and $\mathbb C$ denote the real space and the complex space, respectively. The boldface upper (lower) case letters are used to denote matrices (vectors). ${\bm I}_N$ stands for the $N \times N$ dimensional identity matrix and $\bm 0$ denotes the all-zero vector or matrix. ``~$\setminus$ " represents the set subtraction operation. Superscript $(\cdot)^H$ denotes the conjugated-transpose operation and ${\mathbb E}\{\cdot\}$ denotes the expectation operation. $\left\| {\bm g} \right\|$ is used to give the Euclidean norm of $\bm g$.
\section{System Model}
\subsection{Signal Transmission Model}
\begin{figure}
  \centering
  \includegraphics[scale=0.58]{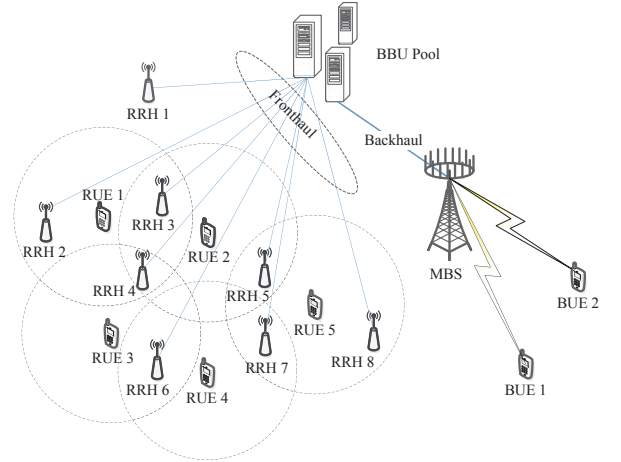}
  \caption{Illustration of an H-CRAN.}
  \label{Fig.1}
\end{figure}

Consider the downlink of a dense H-CRAN with an MBS, a BBU pool, $K$ RRHs and $M$ UEs as shown in Fig. \ref{Fig.1}, where each RRH connects with the BBU pool through an optical fiber and the MBS connects with the BBU pool through a backhaul link. The MBS and each RRH are respectively equipped with $B$ and $N$ antennas, and each UE has a single antenna. Denote the sets of RRHs and UEs by $\cal K$ and $\cal M$, respectively. As discussed in \cite{dai2014sparse}, there are usually two types of clustering methods for RRHs to serve UEs namely disjoint clustering and user-centric clustering. In this paper, user-centric cluster method is adopted, i.e., each UE prefers to access the network via a selected subset of neighboring RRHs and different clusters for different UEs may overlap. If a UE cannot be served by any RRH, the MBS will offer network access to guarantee seamless coverage. For example, in Fig. \ref{Fig.1}, each RUE is served by RRHs inside the circle centered on this RUE. Hence, RUE 1, 2, 3, 4, 5 are served by RRHs, and BUE 1 and BUE 2 are served by the MBS. Denote the sets of RUEs and BUEs by ${\cal M}_{\text R}$ and ${\cal M}_{\text B}$, respectively. Denote ${\cal K}_i \subseteq {\cal K}$ and ${\cal M}_k \subseteq {\cal M}_{\text R}$ as the set of RRHs serving RUE $i$ and the set of RUEs served by RRH $k$, respectively. Assume that all transmitters use the same time-frequency RB to transmit signals, leading to cochannel interference. Then, the received signal at UE $m$ is given by
\begin{equation}
y_m = \sum\limits_{i \in {\cal M}_{\text R}} \sum\limits_{k \in {\cal K}_i} \bm h_{k,m}^H \bm w_{k,i} x_i + \sum\limits_{j \in {\cal M}_{\text B}} \bm h_{b,m}^H \bm w_{b,j} x_{j} +  n_m,
\label{signal_UE}
\end{equation}
where $\bm h_{k,m} \in {\mathbb C}^{N\times1}$ represents the channel vector from RRH $k$ to UE $m$, $\bm w_{k,i} \in {\mathbb C}^{N\times1}$ denotes the beam-vector adopted by RRH $k$ for transmitting signal to RUE $i$, and $x_i$ is the zero-mean unit-variance data symbol for RUE $i$. Likewise, $\bm h_{b,m}, \bm w_{b,j} \in {\mathbb C}^{B\times1}$ and $x_{j}$ are similarly defined for the MBS and BUE $j$, and $n_m$ is the complex white Gaussian noise with variance $N_0$, i.e., $n_m \sim {\cal CN}(0,N_0)$.

\subsection{Channel Estimation}
In most of the related literature, it is usually assumed that the global CSI over the network is available for optimization. To this end, orthogonal pilots should be adopted for channel estimation. However, in a dense H-CRAN network with large numbers of UEs, obtaining CSI of all links is almost infeasible due to limited training resources. As a result, in this paper, orthogonal pilots are adopted by BUEs and RUEs in the same RRH cluster for channel estimation, while pilot reuse among RUEs in different RRH clusters is allowed to reduce pilot overhead. For further pilot overhead reduction, the pilot sequence used by a BUE is allowed to be shared with RUEs. For example, in Fig. \ref{Fig.1}, BUE 1 and BUE 2 both access the network via the MBS. Hence, BUE 1 and BUE 2 adopt orthogonal training for channel estimation. RUE 1 and RUE 2 are both served by RRH 3 and RRH 4. Therefore, RUE 1 and RUE 2 belong to the same cluster and they should be assigned orthogonal pilots. Since the set of RRHs serving RUE 1 and the set of RRHs serving RUE 5 do not overlap, RUE 1 could reuse the pilot of RUE 5. In addition, due to the long distance between RUE 1 and BUE 2, they may be allocated the same pilot for pilot overhead reduction.

In the uplink training phase, assume that RRH $k$ estimates the CSI from all RUEs in ${\cal M}_k$ and the MBS estimates the CSI from all BUEs. The large-scale channel gains from all UEs to each RRH and to the MBS, i.e., $\left\{ \alpha_{k,m}, ~\forall~ k \in {\cal K},~ m \in {\cal M}\right\}$ and $\left\{ \alpha_{b,m}, ~\forall~ m \in {\cal M}\right\}$, are assumed to be available at both the BBU pool and the MBS. Denote ${\cal Q}=\{1,\cdots,\tau\}$ as the available pilot set and ${\bm Q} = \left[{\bm q}_1,\cdots,{\bm q}_{\tau}\right] \in {\mathbb C}^{\tau \times \tau}$ as the pilot matrix with orthogonal column vectors (i.e., ${\bm Q}^H{\bm Q} = {\bm I}_{\tau}$). $\tau$ ($\left| {\cal M}_{\text B}\right| \leq \tau \leq M$) is the length of the pilots and is also the number of pilots available for channel estimation (this is the smallest amount of pilots that are required). Let ${\cal P}({\cal M},{\cal Q})=\left\{(m,\pi_m)|~ m\in {\cal M},~ \pi_m \in {\cal Q}\right\}$ denote an arbitrary pilot assignment scheme, where $(m,\pi_m)$ means that UE $m$ is allocated pilot ${\bm q}_{\pi_m}$. In addition, let ${\cal U}_\pi= \left\{i|~\pi_i=\pi,\right.$ $\left. \forall~ i \in {\cal M}_{\text R} \right\}$ and ${\cal V}_\pi=\left\{j|~\pi_j=\pi, ~\forall~ j \in {\cal M}_{\text B} \right\}$, respectively, denote the sets of RUEs and BUEs that use pilot ${\bm q}_\pi$ for channel estimation. Note that since BUEs apply orthogonal pilots for channel estimation, it follows that $\left|{\cal V}_\pi \right| \in \{0,1\}$.
\subsubsection{Channel Estimation for RUEs}
Given the pilot assignment scheme ${\cal P}({\cal M},{\cal Q})$, the $N\times \tau$ dimensional received signal matrix of pilots at RRH $k$ can be written as
\begin{equation}
{\bm Y}_k^{({\text R})} = \sum\limits_{i \in {\cal M}_{\text R}} {\sqrt {p_{\text R}}} \bm h_{k,i} \bm q_{\pi_i}^H + \sum\limits_{j \in {\cal M}_{\text B}} {\sqrt {p_{\text B}}} \bm h_{k,j} \bm q_{\pi_j}^H +  \bm N_k^{({\text R})},
\label{pilot_RUE}
\end{equation}
where $p_{\text R}$ and $p_{\text B}$ are, respectively, the pilot transmit powers of RUEs and BUEs. $\bm N_k^{({\text R})}$ is the noise matrix which consists of independently and identically distributed (i.i.d.) Gaussian elements with zero mean and variance $N_0$. Then, the minimum mean square error (MMSE) estimate of $\bm h_{k,i}, ~\forall~ k \in {\cal K},~ i \in {\cal M}_k$ is given by \cite{kailath2000linear}
\begin{equation}
{{\hat {\bm h}}_{k,i}} = \frac{\sqrt{p_{\text R}} \alpha_{k,i}}{\sum\limits_{i' \in {\cal U}_{\pi_i}} p_{\text R} \alpha_{k,i'} + \sum\limits_{j \in {\cal V}_{\pi_i}} p_{\text B} \alpha_{k,j} + N_0} {\bm Y}_k^{({\text R})} \bm q_{\pi_i}.
\label{estimation_RUE}
\end{equation}
Given the channel estimate vector ${{\hat {\bm h}}_{k,i}}$, the true channel vector ${\bm h}_{k,i}$ can be expressed as ${\bm h}_{k,i} = {\hat {\bm h}_{k,i}} + {\tilde {\bm h}_{k,i}}$, where the error vector ${\tilde {\bm h}_{k,i}}$ represents the CSI uncertainty. Due to the property of MMSE estimation \cite{kailath2000linear}, ${\tilde {\bm h}_{k,i}}$ is statistically independent of ${\hat {\bm h}_{k,i}}$ and it follows that ${\cal CN}(\bm 0, \delta_{k,i} \bm I_N)$, where $\delta_{k,i}$ is given by
\begin{equation}
\delta_{k,i} = \frac{\alpha_{k,i} \left(\sum\limits_{i' \in {\cal U}_{\pi_i}\setminus i} p_{\text R} \alpha_{k,i'} + \sum\limits_{j \in {\cal V}_{\pi_i}} p_{\text B} \alpha_{k,j} + N_0 \right)}{\sum\limits_{i' \in {\cal U}_{\pi_i}} p_{\text R} \alpha_{k,i'} + \sum\limits_{j \in {\cal V}_{\pi_i}} p_{\text B} \alpha_{k,j} + N_0}.
\label{error_RUE}
\end{equation}
\subsubsection{Channel Estimation for BUEs}
Similarly, the $B \times \tau$ dimensional received signal matrix of pilots at the MBS can be written as
\begin{equation}
{\bm Y}^{({\text B})} = \sum\limits_{i \in {\cal M}_{\text R}} {\sqrt {p_{\text R}}} \bm h_{b,i} \bm q_{\pi_i}^H + \sum\limits_{j \in {\cal M}_{\text B}} {\sqrt {p_{\text B}}} \bm h_{b,j} \bm q_{\pi_j}^H +  \bm N^{({\text B})},
\label{pilot_BUE}
\end{equation}
where $\bm N^{({\text B})}$ is the noise matrix which consists of i.i.d. Gaussian elements with zero mean and variance $N_0$. Then, the MMSE estimate of $\bm h_{b,j}, ~\forall~ j \in {\cal M}_{\text B}$ is
\begin{equation}
{{\hat {\bm h}}_{b,j}} = \frac{\sqrt{p_{\text B}} \alpha_{b,j}}{\sum\limits_{i \in {\cal U}_{\pi_j}} p_{\text R} \alpha_{b,i} + p_{\text B} \alpha_{b,j} + N_0} {\bm Y}^{({\text B})} \bm q_{\pi_j}.
\label{estimation_BUE}
\end{equation}
It follows ${\bm h}_{b,j} = {\hat {\bm h}_{b,j}} + {\tilde {\bm h}_{b,j}}$, and ${\tilde {\bm h}_{b,j}} \sim {\cal CN}(\bm 0, \delta_{b,j} \bm I_B)$ is statistically independent of ${\hat {\bm h}_{b,j}}$, where $\delta_{b,j}$ is given by
\begin{equation}
\delta_{b,j} = \frac{\alpha_{b,j} \left(\sum\limits_{i \in {\cal U}_{\pi_j}} p_{\text R} \alpha_{b,i} + N_0 \right)}{\sum\limits_{i \in {\cal U}_{\pi_j}} p_{\text R} \alpha_{b,i} + p_{\text B} \alpha_{b,j} + N_0}.
\label{error_BUE}
\end{equation}

In general, sum SE is a very important metric in evaluating a wireless network's performance, and sum SE maximization has been widely studied in different kinds of networks \cite{christensen2008weighted,shi2011iteratively,xu2016channel}. Since pilot reuse is assumed in this paper, pilot contamination inevitably exists. Hence, how to effectively mitigate pilot contamination is also important. A two-stage optimization framework is thus studied in the following for network performance maximization. Specifically, the sum MSE of channel estimation is minimized by designing a pilot scheduling algorithm in Section III. Based on the pilot allocation result obtained in Section III, the sum SE of the network is then maximized by optimizing beam-vectors under imperfect CSI in Section IV.
\section{Stage I: Pilot Scheduling}
In this stage, a pilot scheduling algorithm is designed to allocate pilots to UEs based on the metric of minimizing the sum MSE of channel estimation.
\subsection{Problem Formulation}
Since pilots are reused among UEs to shorten pilot overhead, pilot contamination inevitably exists. Considering the location dispersion of UEs, it is preferred that pilot contamination can be effectively mitigated by designing an appropriate pilot scheduling algorithm. According to (\ref{error_RUE}) and (\ref{error_BUE}), the sum MSE of all channel CSI is given by
{\setlength\arraycolsep{2pt}%用于缩短等号两边的空格
\begin{eqnarray}
&&\sum \limits_{i \in {\cal M}_{\text R}} \sum\limits_{k \in {\cal K}_i} {\mathbb E} \left\{ \left\|{\tilde {\bm h}_{k,i}}\right\|^2\right\} + \sum \limits_{j \in {\cal M}_{\text B}} {\mathbb E} \left\{ \left\|{\tilde {\bm h}_{b,j}}\right\|^2\right\}\nonumber\\
&&= \sum \limits_{i \in {\cal M}_{\text R}} \sum\limits_{k \in {\cal K}_i} N \delta_{k,i} + \sum \limits_{j \in {\cal M}_{\text B}} B \delta_{b,j}.
\label{sum_MSE}
\end{eqnarray}}
\!\!To distinguish the channels from different RUEs, it is assumed that any two RUEs for which the sets of serving RRHs are (partially) overlapping would need to be allocated orthogonal pilots. This can be mathematically expressed as follows
\begin{equation}
\pi_i\neq \pi_{i'}, ~\forall~ i,~ i' \in {\cal M}_{\text R},~ i\neq i',~ {\cal K}_i \cap {\cal K}_{i'}\neq \emptyset.
\label{orthogonal_constraint}
\end{equation}

Thus, the problem of minimizing the sum MSE of channel estimation can be formulated as
{\setlength\arraycolsep{2pt}%用于缩短等号两边的空格
\begin{eqnarray}
\mathop {\min }\limits_{{\cal P}({\cal M},{\cal Q})} && \sum \limits_{i \in {\cal M}_{\text R}} \sum\limits_{k \in {\cal K}_i} N \delta_{k,i} + \sum \limits_{j \in {\cal M}_{\text B}} B \delta_{b,j} \label{pilot_scheduling_a}\nonumber\\
\text{s.t.} \;\;\; &&\; ({\text {\ref{orthogonal_constraint}}}). \label{pilot_scheduling_b}
\label{pilot_scheduling}
\end{eqnarray}}

Problem (\ref{pilot_scheduling}) is a resource allocation problem, which can be readily transformed to an equivalent mixed integer programming problem which is, however, usually difficult to solve. The optimal pilot scheduling scheme can be obtained through exhaustive search (ES). However, the complexity of ES increases exponentially with the number of UEs, becoming infeasible for a dense H-CRAN. Therefore, a low complexity pilot scheduling algorithm is proposed in the following subsection.
\subsection{Pilot Scheduling Algorithm}
\begin{figure}
  \centering
  \includegraphics[scale=0.64]{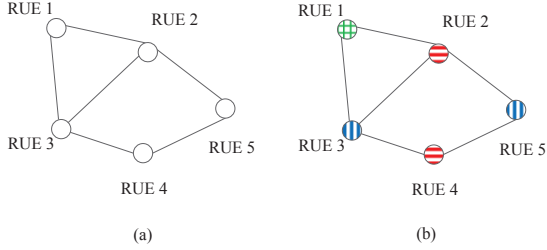}
  \caption{(a) Undirected graph consisting of all RUEs in Fig. \ref{Fig.1}; (b) The colored undirected graph after applying the Dsatur algorithm.}
  \label{Fig.2}
\end{figure}

Constraint (\ref{orthogonal_constraint}) indicates that any two RUEs served by at least one common RRH should be allocated different pilots. This constraint can be equivalently represented by an $\left| {\cal M}_{\text R} \right| \times \left| {\cal M}_{\text R} \right|$ dimensional matrix $\bm A$ with each element given by
\begin{equation}
a_{i,i'} \!=\! \left\{
\begin{array}{ll}
\!\!\!1,& \!\!\! {\text {if}}\, i\neq i',~ {\cal K}_i \cap {\cal K}_{i'}\neq \emptyset \\
\!\!\!0,& \!\!\! {\text {otherwise}}\\
\end{array} \right.\!\!\!,\, \forall~ i,~ i' \in {\cal M}_{\text R}.
\label{constraint_matrix}
\end{equation}
In matrix $\bm A$, when two RUEs are served by at least one common RRH, the corresponding element is one. Otherwise, the element is zero. Obviously, to satisfy constraint (\ref{orthogonal_constraint}), a minimum number of pilots $t$ exists. In order to obtain $t$, an undirected graph can be constructed to describe constraint (\ref{orthogonal_constraint}) based on $\bm A$, where any two RUEs served by at least one common RRH are connected with each other. Then, determining $t$ is equivalent to coloring the vertices of the undirected graph with the minimum number of different colors, which can be optimally solved by using the Dsatur algorithm proposed in \cite{brelaz1979new}. Note that the pilots used by BUEs are mutually orthogonal. Therefore, the length of the pilots should satisfy $\max \left\{\left| {\cal M}_{\text B}\right|,t \right\} \leq \tau \leq M$. Taking the H-CRAN in Fig. \ref{Fig.1} for example, the undirected graph consisting of all RUEs in Fig. \ref{Fig.1} can be depicted as Fig. \ref{Fig.2} (a), and the colored graph of Fig. \ref{Fig.2} (a) after applying the Dsatur algorithm is shown in Fig. \ref{Fig.2} (b). It can be seen that for the considered case, $t=3$. Since $M=7$ and $\left| {\cal M}_{\text B}\right|=2$, the pilot length should satisfy $3\leq \tau \leq 7$.

Though the Dsatur algorithm can yield the minimum number of pilots required by RUEs as well as a feasible pilot allocation solution, it only takes into account constraint (\ref{orthogonal_constraint}) while ignores the objective function of problem (\ref{pilot_scheduling}). Hence, this solution may not be satisfactory. For example, in Fig. \ref{Fig.2} (b), RUEs in the same color share the same pilot, and RUEs in different colors use orthogonal pilots for channel estimation. There may exist measurable pilot contamination between RUE 2 and RUE 4 since they are not so far away from each other. If $\tau = t$, by exchanging the colors of RUE 2 and RUE 1 without changing $t$, the pilot contamination may be reduced due to the relatively longer distance between RUE 1 and RUE 4. If $\tau > t$, the pilot contamination can be certainly decreased by replacing a reused pilot in Fig. \ref{Fig.2} (b) with an unused one.

Since channel estimation is mainly affected by pilot contamination besides the effect of noise, it is thus of great importance to further mitigate pilot contamination. To this end, it is necessary to adjust the pilots allocated to each RUE by the Dsatur algorithm. As stated above, it is difficult to solve problem (\ref{pilot_scheduling}) in polynomial time. Hence, a heuristic low-complexity pilot scheduling algorithm is provided in the following. In order to implement this algorithm, it is first required to figure out how to measure the level of pilot contamination. In \cite{zhu2015graph}, a multi-cell massive MIMO network was considered and pilot reuse was assumed among different cells. To mitigate pilot contamination, a metric was defined to indicate the interference strength among UEs, and a graph coloring based scheme was then proposed. In \cite{xu2017pilot}, a continuous-valued metric was defined to evaluate the potential interference strength between any two D2D pairs when the same pilot was reused, and a pilot scheduling algorithm was then provided based on this metric. Motivated by these works, a similar metric based on large-scale channel gains is defined to measure the level of pilot contamination between an RUE and another UE when they reuse the same pilot. First, for any two unconnected RUEs in the undirected graph, define
{\setlength\arraycolsep{2pt}%用于缩短等号两边的空格
\begin{eqnarray}
&&\beta_{i,i'}=\ln \left(1+\frac{\sum \limits_{k \in {\cal K}_i} \alpha_{k,i'}}{\sum \limits_{k \in {\cal K}_i} \alpha_{k,i}} + \frac{\sum \limits_{k' \in {\cal K}_{i'}} \alpha_{k',i}}{\sum \limits_{k' \in {\cal K}_{i'}} \alpha_{k',i'}} \right), \nonumber\\
&&\;\;\forall~ i,~ i' \in {\cal M}_{\text R},~ i\neq i',~ {\cal K}_i \cap {\cal K}_{i'}= \emptyset.
\label{metric_RUE_unconnected}
\end{eqnarray}}
\!\!The definition of $\beta_{i,i'}$ is inspired by the channel estimation error in (\ref{error_RUE}). Inside the $\ln (\cdot)$ operation of (\ref{metric_RUE_unconnected}), the second term is defined to measure the level of pilot contamination experienced by RUE $i$ from RUE $i'$ if they are assigned the same pilot, and vice versa for the third term. Obviously, a larger $\beta_{i,i'}$ means more severe potential pilot contamination between RUE $i$ and RUE $i'$ when they are assigned the same pilot. For any RUE $i$, let $\beta_{i,i}=0$. For any two connected RUEs, since they are assigned orthogonal pilots, there will be no potential pilot contamination between them. Hence,
\begin{equation}
\beta_{i,i'}=0,~ \forall~ i,~ i' \in {\cal M}_{\text R},~ i\neq i',~ {\cal K}_i \cap {\cal K}_{i'}\neq \emptyset.
\label{metric_RUE_connected}
\end{equation}
As for an RUE and a BUE, define
\begin{equation}
\beta_{i,j}=\ln \left(1+\frac{\sum \limits_{k \in {\cal K}_i} \alpha_{k,j}}{\sum \limits_{k \in {\cal K}_i} \alpha_{k,i}} + \frac{\alpha_{b,i}}{\alpha_{b,j}} \right), ~\forall~ i \in {\cal M}_{\text R},~ j \in {\cal M}_{\text B}.
\label{metric_RUE_BUE}
\end{equation}

Motivated by \cite{zhu2015graph} and \cite{xu2017pilot}, a pilot scheduling algorithm (PSA) is proposed and summarized in Algorithm \ref{PSA}. According to Algorithm \ref{PSA}, it is necessary to first check whether $\tau$ is larger than or equal to the minimum required number of orthogonal pilots or not. If the number of available pilots are not enough to satisfy the scheduled constraints, $\tau$ has to be increased. Then, without loss of generality, pilot ${\bm q}_j$ is allocated to BUE $j$, and $\left\{{\bm q}_1,\cdots,{\bm q}_t\right\}$ are randomly allocated to $t$ clusters with RUEs in each cluster using the same pilot. In order to further mitigate pilot contamination, three steps are iteratively carried out to adjust the pilots assigned to each RUE. The basic idea is that the RUE experiencing larger pilot contamination possesses a higher priority for pilot adjustment. The main steps in each iteration can be explained as follows. First, RUE $i \in {\cal M}_{\text R}\setminus \Lambda$ experiencing the largest interference is selected. Next, pilot $\pi$ which causes the least interference to RUE $i$ is chosen from the set of available pilots, i.e., ${\cal Q}\setminus {\cal X}_i$. Finally, pilot ${\bm q}_\pi$ is assigned to RUE $i$, and sets ${\cal U}_{\pi_i}$, ${\cal U}_\pi$, $\Lambda$ and ${\cal X}_l, ~\forall~ l \in {\cal M}_{\text R}$ are updated. The algorithm will be carried out for $\left| {\cal M}_{\text R} \right|$ times until all RUEs' pilots have been adjusted.
\begin{algorithm}[h]
\begin{algorithmic}
\caption{Pilot Scheduling Algorithm (PSA)}
\label{PSA}
\STATE \textbf {Initialization:}
\STATE \quad Initialize the set of RUEs which have been allocated pilots, i.e., $\Lambda=\emptyset $.
\STATE \quad Obtain matrix $\bm A$ from (\ref{constraint_matrix}) and calculate $t$ by using the Dsatur algorithm.
\STATE \quad If $\tau < \max \left\{\left| {\cal M}_{\text B}\right|,t \right\}$, set $\tau = \max \left\{\left| {\cal M}_{\text B}\right|,t \right\}$.
\STATE \quad Calculate $\beta_{i,m},~ \forall~ i \in {\cal M}_{\text R},~m \in {\cal M}$.
\STATE \textbf {Pilot Allocation:}
\STATE \quad Assign pilot ${\bm q}_j$ to BUE $j, ~\forall~ j \in {\cal M}_{\text B}$.
\STATE \quad Divide all RUEs into $t$ clusters by using the Dsatur algorithm, and randomly allocate pilots $\left\{{\bm q}_1,\cdots,{\bm q}_t\right\}$ to them with RUEs in each cluster using the same pilot. Let ${\cal X}_i$ denote the set of different pilots allocated to the RUEs connected to RUE $i, ~\forall~ i \in {\cal M}_{\text R}$.
\FOR {$d=1,\cdots, \left|{\cal M}_{\text R}\right|$}
\vspace{0.3em}
\STATE 1: $i= \arg \mathop { \max }\limits_{l \in {\cal M}_{\text R}\setminus \Lambda} \left(\sum\limits_{i' \in {\cal U}_{\pi_l}}\! \beta_{l,i'} + \sum\limits_{j \in {\cal V}_{\pi_l}} \beta_{l,j}\right)$.
\vspace{0.3em}
\STATE 2: $\pi=\arg \mathop { \min }\limits_{\iota \in {\cal Q}\setminus {\cal X}_i} \left(\sum\limits_{i' \in {\cal U}_\iota}\! \beta_{i,i'} + \sum\limits_{j \in {\cal V}_\iota} \beta_{i,j}\right)$.
\vspace{0.3em}
\STATE 3: ${\cal U}_{\pi_i}={\cal U}_{\pi_i} \setminus i$, ${\bm q}_{\pi_i}= {\bm q}_\pi$, ${\cal U}_\pi={\cal U}_\pi \cup i$, $\Lambda=\Lambda\cup i$.
\STATE \;\; Update ${\cal X}_l, ~\forall~ l \in {\cal M}_{\text R}$.
\vspace{0.3em}
\ENDFOR
\end{algorithmic}
\end{algorithm}

\subsection{Complexity Analysis}
In this subsection, the computational complexity of Algorithm \ref{PSA} is analyzed with order notation. According to \cite{chen2016training}, the Dsatur algorithm involves a complexity of ${\cal O}\left( \left|{\cal M}_{\text R} \right|^2\right)$. In order to adjust the pilots allocated to RUEs, $\left|{\cal M}_{\text R} \right|$ iterations are carried out. In each iteration, a complexity of ${\cal O}\left( \left|{\cal M}_{\text R} \right|^2\right)$ is required to find the RUE experiencing the largest interference, and the pilot causing the least interference to this RUE. The total complexity of the iteration process is thus ${\cal O}\left( \left|{\cal M}_{\text R} \right|^3\right)$. As a result, Algorithm \ref{PSA} involves an overall complexity of ${\cal O}\left( \left|{\cal M}_{\text R} \right|^3\right)$. In contrast, to obtain the optimal pilot allocation solution, the ES scheme requires a complexity of ${\cal O}\left( \tau^M\right)$, which increases exponentially with the number of UEs. Hence, the proposed algorithm is more efficient for practical use.

\section{Stage II: Robust Transmission Design}
After obtaining the pilot allocation result by using Algorithm \ref{PSA}, the sum SE maximization problem under incomplete inter-cluster CSI and imperfect intra-cluster CSI is considered in this section. A robust transmission design algorithm and its distributed realization are proposed to solve the problem.
\subsection{Problem Formulation}
In order to simplify the expression of (\ref{signal_UE}), the beam-vectors from all RRHs in set ${\cal K}_i$ for transmitting signal to RUE $i$ can be merged to form a large-dimension vector $\bm w_i$, i.e., $\bm w_i = \left[ \bm w_{k,i}^H, ~\forall~ k \in {\cal K}_i \right]^H \in {\mathbb C}^{N|{\cal K}_i| \times 1}$. Similarly, let $\bm g_{i,m} = \left[ \bm h_{k,m}^H, ~\forall~ k \in {\cal K}_i \right]^H \in {\mathbb C}^{N|{\cal K}_i| \times 1}$ represent the aggregated channel vector from all RRHs in ${\cal K}_i$ to UE $m$. Then, (\ref{signal_UE}) can be reformulated as
\begin{equation}
y_m = \sum\limits_{i \in {\cal M}_{\text R}} \bm g_{i,m}^H \bm w_i x_i + \sum\limits_{j \in {\cal M}_{\text B}} \bm h_{b,m}^H \bm w_{b,j} x_j +  n_m.
\label{signal_UE1}
\end{equation}

In the following, all transmit beam-vectors are designed based on the obtained channel estimates, i.e., (\ref{estimation_RUE}), (\ref{error_RUE}), (\ref{estimation_BUE}) and (\ref{error_BUE}), and channel statistics, i.e., the large-scale channel gains. Consider the block fading model, where all channels remain unchanged over the coherence interval with length $T$. Then, the effective SINR and the achievable rate of RUE $i$ are, respectively, given by
{\setlength\arraycolsep{2pt}%用于缩短等号两边的空格
\begin{eqnarray}
\!\!\!\!\!&&\eta_i^{(\text R)} \!=\! \frac{\left|{\hat {\bm g}}_{i,i}^H \bm w_i\right|^2}{\left|{\tilde {\bm g}}_{i,i}^H \bm w_i\right|^2 \!+\! \sum\limits_{i' \in {\cal M}_{\text R}\setminus i}\! \left|\bm g_{i',i} \bm w_{i'}\right|^2 \!+\! \sum\limits_{j \in {\cal M}_{\text B}}\! \left|\bm h_{b,i}^H \bm w_{b,j}\right|^2 \!+ \! N_0},\nonumber\\
\!\!\!\!\!&&R_i^{(\text R)} = \frac{T-\tau}{T} {\mathbb E}\left\{ \log_2 \left(1+ \eta_i^{(\text R)}\right)\right\}, ~\forall~ i \in {\cal M}_{\text R},
\label{rate_RUE}
\end{eqnarray}}
\!\!\!\!where ${\hat {\bm g}}_{i,i} = \left[ {\hat {\bm h}}_{k,i}^H, ~\forall~ k \in {\cal K}_i \right]^H \in {\mathbb C}^{N|{\cal K}_i| \times 1}$ and ${\tilde {\bm g}}_{i,i} = \left[ {\tilde {\bm h}}_{k,i}^H, ~\forall~ k \in {\cal K}_i \right]^H \in {\mathbb C}^{N|{\cal K}_i| \times 1}$, respectively, denote the aggregated channel estimation vector and the aggregated error vector from all RRHs in ${\cal K}_i$ to RUE $i$. Since only MMSE estimates of the channel vectors and the distribution of channel estimation error are available, as in \cite{pan2017jointJSAC2} and \cite{xu2017pilot}, the useful signal in (\ref{rate_RUE}) only contains $\left|{\hat {\bm g}}_{i,i}^H \bm w_i\right|^2$, and the terms corresponding to the channel estimation errors are regarded as Gaussian noise. The expectation operation in (\ref{rate_RUE}) is taken over the unknown channel estimation errors ${\tilde {\bm h}}_{k,i}, ~\forall~ k \in {\cal K}_i$, the inter-cluster channel vectors $\bm h_{k,i}, ~\forall~ k \in {\cal K}\setminus {\cal K}_i$ and $\bm h_{b,i}$.

Similarly, the effective SINR and the achievable rate of BUE $j$ can be, respectively, written as
{\setlength\arraycolsep{2pt}%用于缩短等号两边的空格
\begin{eqnarray}
\!\!\!\!\!&&\eta_j^{(\text B)} \!=\! \frac{\left|\!{\hat {\bm h}}_{b,j}^H \bm w_{b,j}\!\right|^2}{\left|{\tilde {\bm h}}_{b,j}^H \bm w_{b,j}\right|^2 \!\!\!+\!\! \sum\limits_{i \in {\cal M}_{\text R}}\!\! \left|\bm g_{i,j} \bm w_i\right|^2 \!+\! \sum\limits_{j' \in {\cal M}_{\text B}\setminus j}\! \left|\bm h_{b,j}^H \bm w_{b,j'}\right|^2 \!\!+\! \! N_0},\nonumber\\
\!\!\!\!\!&&R_j^{(\text B)} = \frac{T-\tau}{T} {\mathbb E}\left\{ \log_2 \left(1+ \eta_j^{(\text B)}\right)\right\}, ~\forall~ j \in {\cal M}_{\text B},
\label{rate_BUE}
\end{eqnarray}}
\!\!\!\!where the expectation is taken over the unknown channel estimation error ${\tilde {\bm h}}_{b,j}$ and $\bm h_{k,j}, ~\forall~ k \in {\cal K}$.

Due to the fractional form of the SINR expressions and the $\log(\cdot)$ operation, it is difficult to obtain explicit expressions of the achievable rate. In the following theorem, a lower bound on the achievable rate is derived.
\begin{theorem}
Given the SINR formulas in (\ref{rate_RUE}) and (\ref{rate_BUE}), the achievable rates $R_i^{(\text R)}$ and $R_j^{(\text B)}$ are, respectively, lower bounded by (\ref{rate_lb_RUE}) and (\ref{rate_lb_BUE}) as follows
{\setlength\arraycolsep{2pt}% 用于缩短等号两边的空格
\begin{eqnarray}
&& r_i^{(\text R)} = \frac{T-\tau}{T} \log_2 \left(1+ \frac{\left|{\hat {\bm g}}_{i,i}^H \bm w_i\right|^2}{J_i^{(\text R)}}\right), ~\forall~ i \in {\cal M}_{\text R},\label{rate_lb_RUE} \\
&& r_j^{(\text B)} = \frac{T-\tau}{T} \log_2 \left(1+ \frac{\left|{\hat {\bm h}}_{b,j}^H \bm w_{b,j}\right|^2}{J_j^{(\text B)}}\right), ~\forall~ j \in {\cal M}_{\text B},\quad\;\; \label{rate_lb_BUE}
\end{eqnarray}}
\!\!\!where
{\setlength\arraycolsep{2pt}%用于缩短等号两边的空格
\begin{eqnarray}
\!\!\!\!\!\!J_i^{(\text R)} =&& \bm w_i^H \bm E_{i,i}^{(\text R)} \bm w_i + \sum\limits_{i' \in {\cal M}_{\text R}\setminus i} \bm w_{i'}^H \bm G_{i',i}^{(\text R)} \bm w_{i'}\nonumber\\
\!\!\!\!\!\!&& + \sum\limits_{j \in {\cal M}_{\text B}} \bm w_{b,j}^H \bm H_{b,i}^{(\text R)} \bm w_{b,j} +  N_0, ~\forall~ i \in {\cal M}_{\text R},\label{inter_noise_RUE}\\
\!\!\!\!\!\!J_j^{(\text B)} =&& \bm w_{b,j}^H \bm E_{b,j}^{(\text B)} \bm w_{b,j} + \sum\limits_{i \in {\cal M}_{\text R}} \bm w_i^H \bm G_{i,j}^{(\text B)} \bm w_i \nonumber\\
\!\!\!\!\!\!&& + \sum\limits_{j' \in {\cal M}_{\text B}\setminus j} \bm w_{b,j'}^H \bm H_{b,j}^{(\text B)} \bm w_{b,j'} +  N_0, ~\forall~ j \in {\cal M}_{\text B}.\label{inter_noise_BUE}
\end{eqnarray}}
\!\!\!$\bm E_{i,i}^{(\text R)}$, $\bm G_{i',i}^{(\text R)}$, $\bm H_{b,i}^{(\text R)}$, $\bm E_{b,j}^{(\text B)}$, $\bm G_{i,j}^{(\text B)}$ and $\bm H_{b,j}^{(\text B)}$ in (\ref{inter_noise_RUE}) and (\ref{inter_noise_BUE}) are all positive definite matrices defined in Appendix \ref{Appendix_A}.
\label{theorem1}
\end{theorem}
\itshape \textbf{Proof:}  \upshape
See Appendix \ref{Appendix_A}.
\hfill $\Box$

This paper aims to maximize the sum SE of the network by designing robust transmit beam-vectors under incomplete inter-cluster CSI and imperfect intra-cluster CSI. As discussed above, explicit expressions of achievable rates of both RUEs and BUEs are unavailable, while their lower bounds can be obtained according to Theorem \ref{theorem1}. The tightness between achievable rates and their lower bounds is verified in the simulation part. Hence, $R_i^{(\text R)}$ and $R_j^{(\text B)}$ are, respectively, replaced with $r_i^{(\text R)}$ and $r_j^{(\text B)}$, and the problem is formulated as follows
{\setlength\arraycolsep{2pt}% 用于缩短等号两边的空格
\begin{subequations}
\begin{align}
\mathop {\max }\limits_{\bm w}  \quad& \sum\limits_{i \in {\cal M}_{\text R}} r_i^{(\text R)} + \sum\limits_{j \in {\cal M}_{\text B}} r_j^{(\text B)} \label{optimize_problem_a}\\
\text{s.t.} \quad\; & \sum\limits_{i \in {\cal M}_k} \left\| \bm w_{k,i} \right\|^2 \leq P_k^{(\text R)}, ~\forall~ k\in {\cal K},\label{optimize_problem_b}\\
& \sum\limits_{j \in {\cal M}_{\text B}} \left\| \bm w_{b,j} \right\|^2 \leq P^{(\text B)},\label{optimize_problem_c}
\end{align}
\label{optimize_problem}
\end{subequations}}
\!\!\!\!\!\!\!where $\bm w$ is the collection of all beam-vectors, including $\bm w_{k,i}, ~\forall~ k\in {\cal K}, ~i \in {\cal M}_k$ and $\bm w_{b,j}, ~\forall~ j \in {\cal M}_{\text B}$, and $P_k^{(\text R)}$ and $P^{(\text B)}$, respectively, denote the maximum transmit power of RRH $k$ and the MBS.
\subsection{Robust Transmission Design}
Problem (\ref{optimize_problem}) has a form similar to the conventional sum SE maximization problems in downlink MISO systems, which can be effectively solved by adopting the weighted minimum mean square error (WMMSE) algorithm based on the following lemma \cite{christensen2008weighted, shi2011iteratively, xu2015robust}. For brevity, the proof of Lemma \ref{lemma1} is omitted.
\begin{lemma}
In a downlink interfering network, if MMSE receive filters are adopted for signal detection, the following relationship between the MMSE and the SINR of each link holds
\begin{equation}
{\text {MMSE}}=\frac{1}{1+{\text {SINR}}}.
\label{MMSE_SINR}
\end{equation}
\label{lemma1}
\end{lemma}

However, two factors make it difficult to directly exploit the WMMSE algorithm to solve problem (\ref{optimize_problem}). First, due to imperfect channel estimation, self-interference exists in the achievable rate expressions, i.e., terms $\bm w_i^H \bm E_{i,i}^{(\text R)} \bm w_i$ and $\bm w_{b,j}^H \bm E_{b,j}^{(\text B)} \bm w_{b,j}$ exist in $r_i^{(\text R)}$ and $r_j^{(\text B)}$, which is different from the typical rate expression under perfect CSI; Second, due to the expectation operation, $\bm E_{i,i}^{(\text R)}$, $\bm G_{i',i}^{(\text R)}$, $\bm H_{b,i}^{(\text R)}$, $\bm E_{b,j}^{(\text B)}$, $\bm G_{i,j}^{(\text B)}$ and $\bm H_{b,j}^{(\text B)}$ are all positive definite matrices rather than conjugate symmetric rank-one matrices as in popular rate expressions of a downlink MISO system. Because of these two factors, Lemma \ref{lemma1} cannot be directly applied. To deal with this difficulty, the following theorem is first introduced.
\begin{theorem}
$r_i^{(\text R)}$ in (\ref{rate_lb_RUE}) can be regarded as the rate of a mobile user in an equivalent downlink interfering MISO network. The MSE and single-tap MMSE receive equalizer of this user are, respectively, given by
{\setlength\arraycolsep{2pt}%用于缩短等号两边的空格
\begin{eqnarray}
&&{\text {MSE}}_i^{(\text R)} = \left|\left(f_i^{(\text R)} \right)^H {\hat {\bm g}}_{i,i}^H \bm w_i - 1\right|^2 + \left|f_i^{(\text R)} \right|^2 J_i^{(\text R)},\quad\label{MMSE_RUE}\\
&&f_i^{(\text R)} = \frac{{\hat {\bm g}}_{i,i}^H \bm w_i}{\bm w_i^H {\hat {\bm g}}_{i,i} {\hat {\bm g}}_{i,i}^H \bm w_i + J_i^{(\text R)}}, ~\forall~ i \in {\cal M}_{\text R}.\label{f_RUE}
\end{eqnarray}}
\!\!Similarly, $r_j^{(\text B)}$ can also be regarded as the rate of a mobile user in an equivalent downlink interfering MISO network, and its MSE and single-tap MMSE receive equalizer are, respectively, given by
{\setlength\arraycolsep{2pt}%用于缩短等号两边的空格
\begin{eqnarray}
&&{\text {MSE}}_j^{(\text B)} = \left|\left(f_j^{(\text B)} \right)^H {\hat {\bm h}}_{b,j}^H \bm w_{b,j} - 1\right|^2 + \left|f_j^{(\text B)} \right|^2 J_j^{(\text B)},\quad\;\label{MMSE_BUE}\\
&&f_j^{(\text B)} = \frac{{\hat {\bm h}}_{b,j}^H \bm w_{b,j}}{\bm w_{b,j}^H {\hat {\bm h}}_{b,j} {\hat {\bm h}}_{b,j}^H \bm w_{b,j} + J_j^{(\text B)}}, ~\forall~ j \in {\cal M}_{\text B}.\label{f_BUE}
\end{eqnarray}}
\label{theorem2}
\end{theorem}
\itshape \textbf{Proof:}  \upshape
See Appendix \ref{Appendix_B}.
\hfill $\Box$

According to Lemma \ref{lemma1} and Theorem \ref{theorem2}, $r_i^{(\text R)}$ and $r_j^{(\text B)}$ can be rewritten as
{\setlength\arraycolsep{2pt}% 用于缩短等号两边的空格
\begin{eqnarray}
r_i^{(\text R)} &=& - \frac{T-\tau}{T} \log_2 {\text {MMSE}}_i^{(\text R)} \nonumber\\
 &=& - \frac{T-\tau}{T} \mathop {\min }\limits_{f_i^{(\text R)}} \log_2 {\text {MSE}}_i^{(\text R)}, ~\forall~ i \in {\cal M}_{\text R},\label{rate_lb_RUE1} \\
r_j^{(\text B)} &=& - \frac{T-\tau}{T} \log_2 {\text {MMSE}}_j^{(\text B)} \nonumber\\
 &=& - \frac{T-\tau}{T} \mathop {\min }\limits_{f_j^{(\text B)}} \log_2 {\text {MSE}}_j^{(\text B)}, ~\forall~ j \in {\cal M}_{\text B},\label{rate_lb_BUE1}
\end{eqnarray}}
\!\!\!and problem (\ref{optimize_problem}) can be equivalently reformulated as \footnote{Note that in the objective function of problem (\ref{optimize_problem1}), $\frac{T-\tau}{T}$ is omitted and $\log(\cdot)$ is replaced with $\ln(\cdot)$ for the convenience of the following analysis.}
{\setlength\arraycolsep{2pt}% 用于缩短等号两边的空格
\begin{eqnarray}
\mathop {\min }\limits_{{\bm w}, {\bm f}} && \sum\limits_{i \in {\cal M}_{\text R}} \ln {\text {MSE}}_i^{(\text R)} + \sum\limits_{j \in {\cal M}_{\text B}} \ln {\text {MSE}}_j^{(\text B)} \nonumber\\
\text{s.t.} \; && {\text {(\ref{optimize_problem_b})}},~ {\text {(\ref{optimize_problem_c})}},
\label{optimize_problem1}
\end{eqnarray}}
\!\!\!where $\bm f = \left(f_1^{(\text R)}, \cdots, f_{|{\cal M}_{\text R}|}^{(\text R)}, f_1^{(\text B)}, \cdots, f_{|{\cal M}_{\text B}|}^{(\text B)}\right)^T$.

Though the fractional SINR expressions have been avoided, problem (\ref{optimize_problem1}) is still nonconvex and is generally difficult to solve. To make it tractable, the following auxiliary functions are introduced to remove the $\ln(\cdot)$ operation in (\ref{optimize_problem1})
{\setlength\arraycolsep{2pt}% 用于缩短等号两边的空格
\begin{eqnarray}
S_i^{(\text R)} \left(u_i^{({\text R})}\right) &=& \exp \left(u_i^{({\text R})} - 1\right){\text {MSE}}_i^{({\text R})} - u_i^{({\text R})}, ~\forall~ i \in {\cal M}_{\text R},\nonumber\\
S_j^{(\text B)} \left(u_j^{({\text B})}\right) &=& \exp \left(u_j^{({\text B})} - 1\right){\text {MSE}}_j^{({\text B})} - u_j^{({\text B})}, ~\forall~ j \in {\cal M}_{\text B},\nonumber\\
\label{auxiliary_function}
\end{eqnarray}}
where $u_i^{({\text R})}$ and $u_j^{({\text B})}$ are newly introduced auxiliary variables. Checking the first-order optimality condition of (\ref{auxiliary_function}) yields
{\setlength\arraycolsep{2pt}% 用于缩短等号两边的空格
\begin{eqnarray}
&&\mathop {\min }\limits_{u_i^{({\text R})}} S_i^{(\text R)} \left(u_i^{({\text R})}\right) =  \ln {\text {MSE}}_i^{({\text R})}, ~\forall~ i \in {\cal M}_{\text R},\nonumber\\
&&\mathop {\min }\limits_{u_j^{({\text B})}} S_j^{(\text B)} \left(u_j^{({\text B})}\right) =  \ln {\text {MSE}}_j^{({\text B})}, ~\forall~ j \in {\cal M}_{\text B},
\label{first-order}
\end{eqnarray}}
\!and the corresponding optimal solutions
{\setlength\arraycolsep{2pt}% 用于缩短等号两边的空格
\begin{eqnarray}
&& u_i^{({\text R})*}= 1- \ln {\text {MSE}}_i^{({\text R})}, ~\forall~ i \in {\cal M}_{\text R},\nonumber\\
&& u_j^{({\text B})*}= 1- \ln {\text {MSE}}_j^{({\text B})}, ~\forall~ j \in {\cal M}_{\text B}.
\label{optimal_u}
\end{eqnarray}}

Therefore, according to (\ref{first-order}) and by substituting (\ref{auxiliary_function}) into (\ref{optimize_problem1}), the problem becomes
{\setlength\arraycolsep{2pt}% 用于缩短等号两边的空格
\begin{eqnarray}
\mathop {\min }\limits_{{\bm w}, {\bm f}, {\bm u}} && \sum\limits_{i \in {\cal M}_{\text R}} \left[ \exp \left(u_i^{({\text R})} - 1\right){\text {MSE}}_i^{({\text R})} - u_i^{({\text R})} \right] \nonumber\\
&& + \sum\limits_{j \in {\cal M}_{\text B}} \left[ \exp \left(u_j^{({\text B})} - 1\right){\text {MSE}}_j^{({\text B})} - u_j^{({\text B})} \right] \nonumber\\
\text{s.t.} \;\;\;\, && {\text {(\ref{optimize_problem_b})}},~ {\text {(\ref{optimize_problem_c})}},
\label{optimize_problem2}
\end{eqnarray}}
\!\!\!where $\bm u = \left(u_1^{(\text R)}, \cdots, u_{|{\cal M}_{\text R}|}^{(\text R)}, u_1^{(\text B)}, \cdots, u_{|{\cal M}_{\text B}|}^{(\text B)}\right)^T$. Compared with problem (\ref{optimize_problem1}), problem (\ref{optimize_problem2}) is much easier to solve since it is convex with respect to (w.r.t.) each of the individual variables. By alternatively optimizing $\bm w$, $\bm f$ and $\bm u$, a suboptimal solution of problem (\ref{optimize_problem2}) can be obtained.

For fixed $\bm w$ and $\bm u$, the optimal $\bm f$ can be obtained from (\ref{f_RUE}) and (\ref{f_BUE}). For given $\bm w$ and $\bm f$, the optimal $\bm u$ can be obtained according to (\ref{optimal_u}). When $\bm f$ and $\bm u$ have been determined, for notational brevity, denote $\beta_i^{({\text R})} = \exp \left(u_i^{({\text R})} - 1\right)$, $\beta_j^{({\text B})} = \exp \left(u_j^{({\text B})} - 1\right)$, and delete constants $u_i^{({\text R})}$ as well as $u_j^{({\text B})}$ in the objective function of (\ref{optimize_problem2}). Then, based on (\ref{MMSE_RUE}) and (\ref{MMSE_BUE}), problem (\ref{optimize_problem2}) can be equivalently transformed to \footnote{Note that for brevity, a constant term in the objective function of (\ref{optimize_problem3}) is omitted, which does not affect the equivalence between (\ref{optimize_problem2}) and (\ref{optimize_problem3}).}
{\setlength\arraycolsep{2pt}% 用于缩短等号两边的空格
\begin{eqnarray}
\mathop {\min }\limits_{\bm w} && \sum\limits_{i \in {\cal M}_{\text R}} \left\{\bm w_i^H \bm F_i^{(\text R)} \bm w_i - 2 \beta_i^{(\text R)} {\text {Re}} \left[ \left(f_i^{(\text R)} \right)^H {\hat {\bm g}}_{i,i}^H \bm w_i \right] \right\} \nonumber\\
+&&\!\!\!\! \sum\limits_{j \in {\cal M}_{\text B}} \left\{\bm w_{b,j}^H \bm F_j^{(\text B)} \bm w_{b,j} - 2 \beta_j^{(\text B)} {\text {Re}} \left[ \left(f_j^{(\text B)} \right)^{H} {\hat {\bm h}}_{b,j}^H \bm w_{b,j} \right] \right\}\!\! \nonumber\\
\text{s.t.}  &&\!\! {\text {(\ref{optimize_problem_b})}},~ {\text {(\ref{optimize_problem_c})}},
\label{optimize_problem3}
\end{eqnarray}}
\!\!\!where
{\setlength\arraycolsep{2pt}%用于缩短等号两边的空格
\begin{eqnarray}
\!\!\!&& \bm F_i^{(\text R)} = \beta_i^{(\text R)} \left|f_i^{(\text R)} \right|^2 {\hat {\bm g}}_{i,i} {\hat {\bm g}}_{i,i}^H + \beta_i^{(\text R)} \left|f_i^{(\text R)} \right|^2 \bm E_{i,i}^{(\text R)}\nonumber\\
\!\!\!&& \; + \sum\limits_{i' \in {\cal M}_{\text R}\setminus i} \beta_{i'}^{(\text R)} \left|f_{i'}^{(\text R)} \right|^2 \bm G_{i,i'}^{(\text R)} + \sum\limits_{j \in {\cal M}_{\text B}} \beta_j^{(\text B)} \left|f_j^{(\text B)} \right|^2 \bm G_{i,j}^{(\text B)},\nonumber\\
\!\!\!&& \bm F_j^{(\text B)} = \beta_j^{(\text B)} \left|f_j^{(\text B)} \right|^2 {\hat {\bm h}}_{b,j} {\hat {\bm h}}_{b,j}^H + \beta_j^{(\text B)} \left|f_j^{(\text B)} \right|^2 \bm E_{b,j}^{(\text B)}\nonumber\\
\!\!\!&& \; + \sum\limits_{i \in {\cal M}_{\text R}} \beta_i^{(\text R)} \left|f_i^{(\text R)} \right|^2 \bm H_{b,i}^{(\text R)} + \sum\limits_{j' \in {\cal M}_{\text B}\setminus j} \beta_{j'}^{(\text B)} \left|f_{j'}^{(\text B)} \right|^2 \bm H_{b,j'}^{(\text B)}.\quad\;\;
\label{F}
\end{eqnarray}}
\!\!\!From Theorem \ref{theorem1} and (\ref{F}), it is known that both $\bm F_i^{(\text R)}$ and $\bm F_j^{(\text B)}$ are positive definite matrices. Hence, problem (\ref{optimize_problem3}) is a quadratically constrained quadratic programming (QCQP), and can be optimally solved by adopting a standard convex optimization solver such as CVX, which is a toolbox developed in MATLAB for solving convex problems \cite{grant2008cvx}.

Based on the above analysis, problem (\ref{optimize_problem2}) can be effectively solved by alternatively optimizing ${\bm w}$, ${\bm f}$ and ${\bm u}$. Detailed steps are summarized in Algorithm \ref{RTD}.
\begin{algorithm}[h]
\begin{algorithmic}[1]
\caption{Robust Transmission Design (RTD)}
\label{RTD}
\STATE Set $d=0$, initialize $\bm w(d) = \bm 0$, $\bm f(d) = \bm 1$, $\bm u(d) = \bm 1$ and $\rho=10^{-3}$.
\REPEAT
\STATE Solve QCQP problem (\ref{optimize_problem3}) to obtain $\bm w(d+1)$ by adopting the CVX toolbox.
%\vspace{0.1em}
\STATE Obtain $\bm f(d+1)$ based on (\ref{f_RUE}) and (\ref{f_BUE}).
%\vspace{0.1em}
\STATE Obtain $\bm u(d+1)$ based on (\ref{optimal_u}).
%\vspace{0.1em}
\STATE $d=d+1$.
%\Until $\textcolor[rgb]{0.00,0.07,1.00}{\sum\limits_{m \in {\cal M}_{\text R}} \left\|\bm w_m(d+1) - \bm w_m(d) \right\|^2 +}$\\
% \quad\quad $\textcolor[rgb]{0.00,0.07,1.00}{\sum\limits_{m \in {\cal M}_{\text B}} \left\|\bm w_{b,m}(d+1) - \bm w_{b,m}(d) \right\|^2 \leq \rho.}$
\UNTIL \vspace{-1.6em}
 \begin{align}
\!\!\!\!\!\!&\sum\limits_{i \in {\cal M}_{\text R}} \left\|\bm w_i(d+1) - \bm w_i(d) \right\|^2 +\nonumber\\
\!\!\!\!\!\!&\sum\limits_{j \in {\cal M}_{\text B}} \left\|\bm w_{b,j}(d+1) - \bm w_{b,j}(d) \right\|^2 \leq \rho.
 \nonumber
 \end{align}
\end{algorithmic}
\end{algorithm}
\subsection{Distributed Implementation of the Proposed RTD Algorithm}
\label{IV-C}
Solving problem (\ref{optimize_problem2}) using the proposed RTD algorithm requires a control center with great computation power, especially for an ultra-dense multi-cell network. Hence, it is desirable to obtain the beam-vectors in a decentralized manner with only local CSI, i.e., the BBU pool uses only $\left\{{\hat {\bm h}}_{k,i}, ~\forall~ k \in {\cal K}_i,~i \in {\cal M}_{\text R}\right\}$ to obtain $\bm w^{(\text R)}$, and the MBS uses $\left\{{\hat {\bm h}}_{b,j}, ~\forall~ j \in {\cal M}_{\text B}\right\}$ to obtain $\bm w^{(\text B)}$, where $\bm w^{(\text R)}$ and $\bm w^{(\text B)}$ respectively denote the collections of all beam-vectors used by RRHs and the MBS. Since it is assumed that both the BBU pool and the MBS have global large-scale channel gains, i.e., $\left\{ \alpha_{k,m}, ~\forall~ k \in {\cal K}, m \in {\cal M}\right\}$ and $\left\{ \alpha_{b,m}, ~\forall~ m \in {\cal M}\right\}$, it is shown in the following that the proposed RTD algorithm can, fortunately, be implemented in a distributed manner.

Denote $\bm f^{(\text R)} = \left(f_1^{(\text R)}, \cdots, f_{|{\cal M}_{\text R}|}^{(\text R)}\right)^T$, $\bm f^{(\text B)} = \left(f_1^{(\text B)}, \cdots,\right.$ $\left. f_{|{\cal M}_{\text B}|}^{(\text B)}\right)^T$, $\bm u^{(\text R)} = \left(u_1^{(\text R)}, \cdots, u_{|{\cal M}_{\text R}|}^{(\text R)}\right)^T$ and $\bm u^{(\text B)} = \left(u_1^{(\text B)},\right.$ $\left. \cdots, u_{|{\cal M}_{\text B}|}^{(\text B)}\right)^T$. Then, the $d$th iteration of Algorithm \ref{RTD} can be processed in a decentralized fashion as follows. First, the BBU pool sends $\left(\bm f^{(\text R)} (d-1), \bm u^{(\text R)} (d-1)\right)$ obtained in the $(d-1)$th iteration to the MBS, and the MBS sends $\left(\bm f^{(\text B)} (d-1), \bm u^{(\text B)} (d-1)\right)$ to the BBU pool. Second, divide problem (\ref{optimize_problem3}) into two subproblems with the first subproblem aiming to minimize the first term of the objection function of (\ref{optimize_problem3}) subject to constraint (\ref{optimize_problem_b}) and the second one aiming to minimize the second term of the objection function of (\ref{optimize_problem3}) subject to constraint (\ref{optimize_problem_c}). These two subproblems can be independently solved, and $\bm w^{(\text R)} (d)$ as well as $\bm w^{(\text B)} (d)$ can thus be respectively obtained at the BBU pool and the MBS. The BBU pool and the MBS then exchange the obtained $\bm w^{(\text R)} (d)$ and $\bm w^{(\text B)} (d)$ with each other, and respectively calculate $\left(\bm f^{(\text R)} (d), \bm u^{(\text R)} (d)\right)$ and $\left(\bm f^{(\text B)} (d), \bm u^{(\text B)} (d)\right)$ based on (\ref{f_RUE}), (\ref{f_BUE}) and (\ref{optimal_u}). After checking the termination criterion, the algorithm stops if the algorithm converges. Otherwise, continue to the next iteration.

The computational complexity of executing the proposed RTD algorithm is analyzed in the next subsection, from which it can be seen that, compared with the centralized way, executing the algorithm in a distributed manner can help spread out the compute task of the control center over the BBU pool and the MBS. This will help reduce the computational burden of the control center, especially for a network with multiple cells. The cost of the distributed implementation is the exchange of variables $\left(\bm f^{(\text R)} (d), \bm u^{(\text R)} (d)\right)$, $\bm w^{(\text R)} (d)$, $\left(\bm f^{(\text B)} (d), \bm u^{(\text B)} (d)\right)$ and $\bm w^{(\text B)} (d)$ between the BBU pool and the MBS. However, when distributed implementation is adopted, either the BBU pool or the MBS solves the corresponding problem with only local CSI, i.e., there is no need to collect the overall CSI, which improves the scalability. In addition, it is shown in Section~\ref{simulation} that the RTD algorithm converges rapidly within a few iterations. Hence, the distributed implementation is suitable for practical applications.
\subsection{Convergence and Complexity Analysis}
Since Algorithm \ref{RTD} is carried out in an alternative manner, it is necessary to characterize its convergence behavior. In each iteration, the optimal $\bm w$ is first obtained by solving (\ref{optimize_problem3}). Then, the optimal $\bm f$ and $\bm u$ are obtained according to (\ref{f_RUE}), (\ref{f_BUE}) and (\ref{optimal_u}). As a result, the objective function of (\ref{optimize_problem2}) decreases in each iteration. Due to the fact that this objective function is always lower bounded, the convergence of the proposed RTD algorithm is thus guaranteed.

Then, the computational complexity of the proposed RTD algorithm is analyzed. The complexity of this algorithm mainly lies in solving QCQP problem (\ref{optimize_problem3}). As stated in Subsection \ref{IV-C}, problem (\ref{optimize_problem3}) can be divided into two subproblems, and according to \cite{dai2014sparse}, both of these subproblems can be equivalently transformed to a second-order cone programming (SOCP). The total numbers of variables in the two equivalent SOCP problems are, respectively, $D_1 = \sum\limits_{i \in {\cal M}_{\text R}} N \left|{\cal K}_i \right|$ and $D_2 = B \left|{\cal M}_{\text B} \right|$. Hence, each iteration involves an approximate complexity of ${\cal O} \left( D_1^{3.5}+D_2^{3.5}\right)$ \cite{ye1998interior}. Assume that $L$ iterations are required for Algorithm \ref{RTD} to converge. The total complexity of Algorithm \ref{RTD} is thus ${\cal O} \left( L \left( D_1^{3.5}+D_2^{3.5}\right)\right)$. When the RTD algorithm is carried out in a distributed way, the BBU poll and the MBS will, respectively, solve the two subproblems, and as the centralized way, $L$ iterations are required for the algorithm to converge. Therefore, complexities of ${\cal O} \left( L D_1^{3.5}\right)$ and ${\cal O} \left( L D_2^{3.5}\right)$ are, respectively, involved at the BBU pool and the MBS.
\section{Simulation Results}
\label{simulation}
\begin{table}\centering
\label{Simu_Para}
\caption{Simulation Parameters}
\vspace{-0.8em}
\begin{tabular}{|l|l|}
\hline Radius of the cell & 500 m \\
\hline Pilot power of BUEs $p_{\text B}$& 20 dBm\\
\hline Pilot power of RUEs $p_{\text R}$& 17 dBm\\
\hline Maximum transmit power of the MBS $P^{(\text B)}$& 30 dBm\\
\hline Maximum transmit power of each RRH $P^{(\text R)}$& 27 dBm\\
\hline Additive noise power $N_0$ & -100 dBm\\
\hline Path loss exponent & 3.7\\
\hline Standard deviation of log-normal shadowing fading& 8 dB\\
\hline Accuracy $\rho$ & $10^{-3}$\\
\hline
\end{tabular}
\end{table}
In this section, representative simulation results are presented to evaluate the performance of the proposed algorithms. An isolated H-CRAN is considered with the MBS located at the center of the cell and all UEs randomly distributed. When a UE is close to the MBS, it usually prefers to access the network via the MBS. Hence, it is assumed that all RRHs are uniformly distributed in a ring area centered around the base station with the radius of the inner ring to be $200$ m and the radius of the outer ring equal to the radius of the cell. Each RRH has a covering radius of $D_{\text{max}}$ m. If the distance between a UE and an RRH is within $D_{\text{max}}$ m, the UE chooses to access the network via this RRH. Assume that each RRH can simultaneously serve 3 UEs. If more than 3 UEs are associated with an RRH, the RRH will choose to serve the 3 closest UEs, and the rest UEs will be served by either the other RRHs or the MBS. If a UE cannot be served by any RRHs, the conventional cellular communication serves this UE. For the sake of brevity, equal maximum power constraint for all RRHs is assumed, i.e., $P_k^{(\text R)} = P^{(\text R)}, ~\forall~ k\in {\cal K}$. Unless otherwise specified, the other system parameters are summarized in Table I. All simulation results are obtained by averaging over $1000$ channel realizations, and each channel realization is obtained by generating a random user distribution as well as a random set of fading coefficients.

\begin{figure}
  \centering
  \includegraphics[scale=0.5]{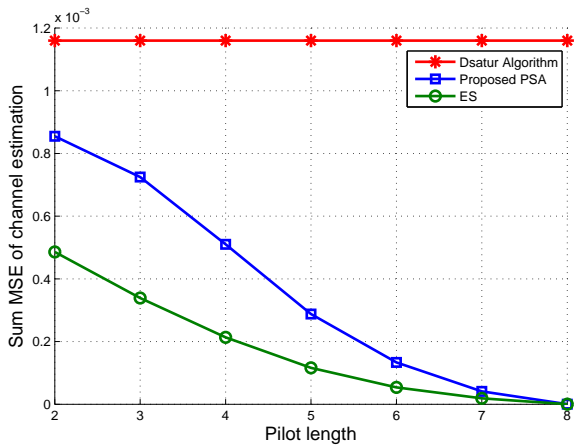}
  \caption{Sum MSE of channel estimation versus pilot length with $M=8$, $K=25$, $B=10$, $N=4$, $D_{\text{max}}=100$ and $T=50$.}
  \label{Fig.3}
\end{figure}
\subsection{Performance of the Proposed PSA alorithm}
In this subsection, the performance of the proposed PSA alorithm is investigated in terms of sum MSE of the network. For comparison, the results obtained by the Dsatur algorithm and the ES scheme are taken as benchmarks. In particular, the Dsatur algorithm divides all RUEs into $t$ clusters, and randomly allocates pilots $\left\{{\bm q}_1,\cdots,{\bm q}_t\right\}$ to RUEs with RUEs in each cluster using the same pilot. The ES scheme searches all feasible pilot allocation schemes and can always find the optimal solution with high calculation cost. Note that in the simulation process, since the UEs and RRHs are randomly generated, if the number of required pilots can not be satisfied for given $\tau$, i.e., $\tau < \max \left\{\left| {\cal M}_{\text B}\right|,t \right\}$, set $\tau = \max \left\{\left| {\cal M}_{\text B}\right|,t \right\}$.

In Fig. \ref{Fig.3}, the sum MSE of channel estimation versus pilot length is depicted. It can be seen from this figure that the proposed PSA algorithm outperforms the Dsatur algorithm greatly in terms of sum MSE. When $\tau$ increases, as expected, the sum MSE obtained by the proposed PSA algorithm decreases with $\tau$ and approaches that obtained by the ES scheme. While for the Dsatur algorithm, the sum MSE remains unchanged since it always assigns $t$ pilots to RUEs regardless of $\tau$.

\begin{figure}
  \centering
  \includegraphics[scale=0.5]{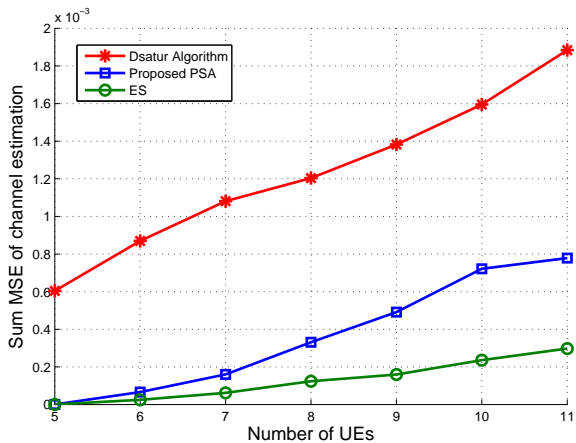}
  \caption{Sum MSE of channel estimation versus the number of UEs with $K=25$, $B=10$, $N=4$, $D_{\text{max}}=100$, $\tau=5$ and $T=50$.}
  \label{Fig.4}
\end{figure}

\begin{figure}
  \centering
  \includegraphics[scale=0.5]{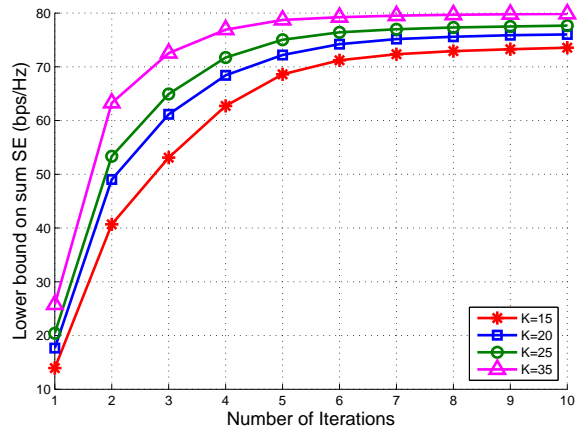}
  \caption{Convergence behaviors of the proposed RTD algorithm with $M=10$, $B=10$, $N=4$, $D_{\text{max}}=100$, $\tau=5$ and $T=50$.}
  \label{Fig.5}
\end{figure}

Fig. \ref{Fig.4} depicts the sum MSE of channel estimation versus the number of UEs. Compared with the Dsatur algorithm, the sum MSE of the network can be decreased greatly by the proposed PSA algorithm. As the number of UEs grows, more channels are required to be estimated and the probability of pilot reusing increases for given $\tau$, leading to more pilot contamination. Hence, the sum MSE increases for all considered cases. When the number of UEs is equal to the pilot length, i.e., $M=\tau$, orthogonal training is the optimal pilot allocation solution. In this case, there will be no pilot contamination, and channel estimation is only affected by thermal noise. Hence, the sum MSE will be very small, which can be found from both Figs. \ref{Fig.3} and \ref{Fig.4}.

\subsection{Performance of the Proposed RTD alorithm}
In this subsection, the performance of the proposed RTD alorithm is evaluated.

First, in Fig. \ref{Fig.5}, the convergence behavior of the proposed RTD algorithm under different values of $K$ is illustrated. It can be seen from this figure that the lower bound on sum SE of the network monotonically increases during the iterative procedure and converges rapidly after only a few iterations (within 8 iterations for all considered configurations). Fig. \ref{Fig.5} also shows that the sum SE increases with the number of RRHs. This is because more UEs will access the network via RRHs and each RUE can be served by more RRHs as $K$ grows. Due to short transmission distance between RUEs and RRHs, SE gains can be obtained in contrast to the conventional cellular communication.

\begin{figure}
  \centering
  \includegraphics[scale=0.5]{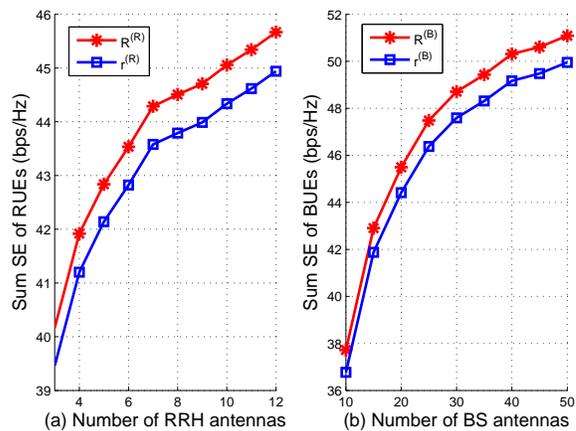}
  \caption{(a) Sum SE of RUEs versus the number of RRH antennas with $M=10$, $K=25$, $B=10$, $D_{\text{max}}=100$, $\tau=5$ and $T=50$; (b) Sum SE of BUEs versus the number of MBS antennas with $M=10$, $K=25$, $N=4$, $D_{\text{max}}=100$, $\tau=5$ and $T=50$.}
  \label{Fig.6}
\end{figure}

Since explicit expressions of achievable rate of both RUEs and BUEs are unavailable, this paper aims to maximize the lower bound on sum SE of the network. It is thus necessary to verify the feasibility. In Fig. \ref{Fig.6}, the gaps between the achievable rate of RUEs, BUEs and their corresponding lower bounds are investigated. Fig. \ref{Fig.6} shows that the achievable sum rates of RUEs and BUEs are both close to their corresponding lower bounds, indicating that it is reasonable to solve the sum SE maximization problem based on lower bounds (\ref{rate_lb_RUE}) and (\ref{rate_lb_BUE}). In addition, Fig. \ref{Fig.6} also shows that the sum SE increases with RRH and MBS antennas, which is consistent with intuition.

\begin{figure}
  \centering
  \includegraphics[scale=0.5]{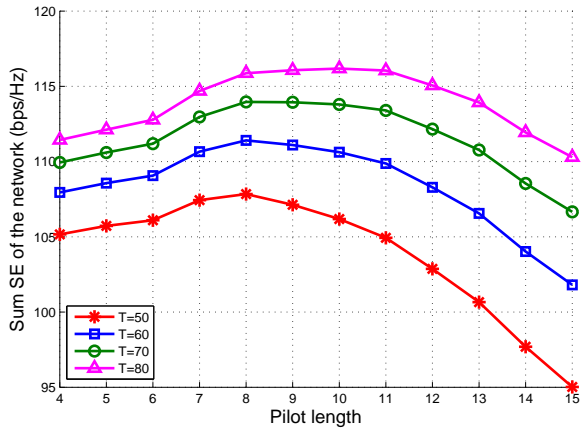}
  \caption{Sum SE of the network versus pilot length with $M=15$, $K=25$, $B=16$, $N=4$ and $D_{\text{max}}=100$.}
  \label{Fig.7}
\end{figure}

\begin{figure}
  \centering
  \includegraphics[scale=0.5]{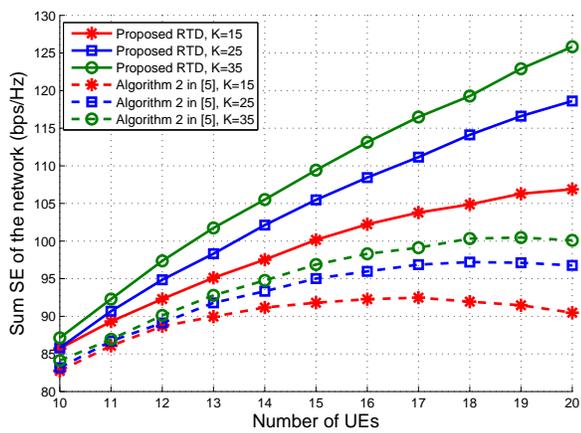}
  \caption{Sum SE of the network versus the number of UEs with $B=16$, $N=4$, $D_{\text{max}}=100$, $\tau=5$ and $T=50$.}
  \label{Fig.8}
\end{figure}

In Fig. \ref{Fig.7}, the effect of pilot length under different values of coherence interval is investigated. It is shown that for all considered cases, the sum SE of the network first grows and then decreases w.r.t. $\tau$. This is because only a few orthogonal pilots are reused among UEs for a small $\tau$. In this case, the channel estimation is significantly influenced by pilot contamination. Therefore, the sum SE can be enhanced by increasing $\tau$. However, as $\tau$ becomes large enough, the channel estimation accuracy can be hardly improved by further enlarging $\tau$. Counterproductively, increasing pilot length reduces the number of symbols available for data transmission, and thereby reduces the sum SE. Note that in Fig. \ref{Fig.7}, the minimum sum SE is obtained when $\tau = M$, i.e., UEs adopt the conventional orthogonal training scheme for channel estimation. Hence, the sum SE of the network can be significantly increased by pilot reuse and the proposed RTD algorithm, especially when $T$ is small.

\begin{figure}
  \centering
  \includegraphics[scale=0.5]{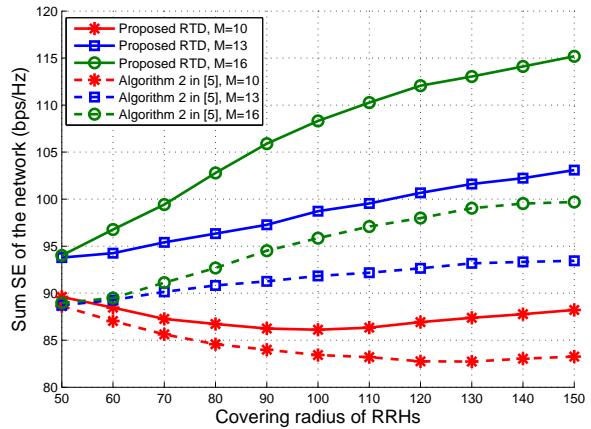}
  \caption{Sum SE of the network versus the covering radius of RRHs with $K=25$, $B=16$, $N=4$, $\tau=5$ and $T=50$.}
  \label{Fig.9}
\end{figure}

Fig. \ref{Fig.8} depicts the sum SE of the network versus the number of UEs under different values of $K$. For comparison, the results obtained by using Algorithm 2 proposed in \cite{dai2014sparse} are depicted as benchmarks, which assume that perfect global CSI is available. Note that reference \cite{dai2014sparse} considered a C-RAN without MBS. Hence, for the sake of fairness, the same H-CRAN network is considered when obtaining the benchmarks. From Fig. \ref{Fig.8}, several observations can be made. First, as expected, the sum SE grows with $K$ for all considered cases. Second, as the number of UEs increases, the sum SE grows monotonically for the proposed RTD algorithm, while first increases and then slightly decreases for Algorithm 2 proposed in \cite{dai2014sparse}. This is because orthogonal training is used by \cite{dai2014sparse} to obtain the perfect CSI. As $M$ grows, the number of required pilots increases, and thereby decreases the sum SE. In addition, it can also be seen that compared with Algorithm 2 proposed in \cite{dai2014sparse}, the sum SE of the network can be significantly improved by the proposed RTD algorithm.
\newcounter{TempEqCnt}
\setcounter{TempEqCnt}{\value{equation}}
\setcounter{equation}{36}
\begin{figure*}[hb]
\hrulefill
{\setlength\arraycolsep{2pt}% 用于缩短等号两边的空格
\begin{eqnarray}
R_i^{(\text R)} &\geq& \frac{T-\tau}{T} \log_2 \left(1+\left({\mathbb E}\left\{\frac{1}{\eta_i^{(\text R)}}\right\}\right)^{-1}\right)\nonumber\\
&=& \frac{T-\tau}{T} \log_2 \left(1+ \frac{\left|{\hat {\bm g}}_{i,i}^H \bm w_i\right|^2}{{\mathbb E}\left\{\left|{\tilde {\bm g}}_{i,i}^H \bm w_i\right|^2\right\} + \sum\limits_{i' \in {\cal M}_{\text R}\setminus i} {\mathbb E}\left\{\left|\bm g_{i',i} \bm w_{i'}\right|^2\right\} + \sum\limits_{j \in {\cal M}_{\text B}} {\mathbb E}\left\{\left|\bm h_{b,i}^H \bm w_{b,j}\right|^2\right\} + N_0}\right)\nonumber\\
&\triangleq& r_i^{(\text R)}, ~\forall~ i \in {\cal M}_{\text R}, \label{Jensen_RUE}
\end{eqnarray}}
\end{figure*}

\setcounter{equation}{42}
\begin{figure*}[hb]
{\setlength\arraycolsep{2pt}% 用于缩短等号两边的空格
\begin{eqnarray}
R_j^{(\text B)} &\geq& \frac{T-\tau}{T} \log_2 \left(1+\left({\mathbb E}\left\{\frac{1}{\eta_j^{(\text B)}}\right\}\right)^{-1}\right)\nonumber\\
&=& \frac{T-\tau}{T} \log_2 \left(1+ \frac{\left|{\hat {\bm h}}_{b,j}^H \bm w_{b,j}\right|^2}{{\mathbb E}\left\{\left|{\tilde {\bm h}}_{b,j}^H \bm w_{b,j}\right|^2\right\} + \sum\limits_{i \in {\cal M}_{\text R}} {\mathbb E}\left\{\left|\bm g_{i,j} \bm w_i\right|^2\right\} + \sum\limits_{j' \in {\cal M}_{\text B}\setminus j} {\mathbb E}\left\{\left|\bm h_{b,j}^H \bm w_{b,j'}\right|^2\right\} + N_0}\right)\nonumber\\
&\triangleq& r_j^{(\text R)}, ~\forall~ j \in {\cal M}_{\text B}, \label{Jensen_BUE}
\end{eqnarray}}
\end{figure*}
In Fig. \ref{Fig.9}, the effect of covering radius of RRHs on sum SE of the network is investigated. As $D_{\text{max}}$ increases, more UEs will access the network via RRHs and each RUE can be served by more RRHs. Hence, for the cases with $M=13$ and $M=16$, the sum SE of the network grows with $D_{\text{max}}$ for both the proposed RTD algorithm and Algorithm 2 proposed in \cite{dai2014sparse}. However, when $M=10$, the sum SE first decreases and then increases for both the proposed RTD algorithm and the benchmark. This is due to the fact that as $D_{\text{max}}$ increases, channel conditions between RRHs and RUEs become worse. Though coverage areas of RRHs become large, the sum SE gains brought by this are limited when $M$ is small. Hence, the sum SE of the network suffers a decrease.

\section{Conclusions}
\label{section6}
This paper has studied pilot scheduling and robust transmission design problems in an ultra-dense H-CRAN. Since pilot reuse was assumed among UEs to shorten pilot overhead, pilot contamination inevitably exists. Hence, a pilot scheduling algorithm was proposed to minimize the sum MSE of all channel estimates. Afterwards, robust transmit beam-vectors were designed to maximize the sum SE of the network. Since each RRH or the MBS only has the imperfect CSI of intra-cluster UEs and only tracks the large-scale channel gains of inter-cluster UEs, it was difficult to obtain the exact achievable rate of each link. Hence, a lower bound on each UE's achievable rate was derived and an alternative robust transmission design algorithm was proposed to maximize the lower bound on sum SE. Simulation results showed that compared with existing algorithms, the system performance can be significantly improved by the proposed algorithms in terms of both sum MSE and sum SE.
\appendices
\section{Proof of Theorem \ref{theorem1}}
\label{Appendix_A}
Using the convexity of $\log_2 \left(1+\frac{1}{x}\right)$ ($\forall~ x >0$) and applying the Jensen's inequality, a lower bound on $R_i^{(\text R)}$ can be derived as (\ref{Jensen_RUE}) at the bottom of this page. Denote
\setcounter{equation}{37}
{\setlength\arraycolsep{2pt}%用于缩短等号两边的空格
\begin{eqnarray}
&&\bm E_{i,i}^{(\text R)}= {\mathbb E}\left\{{\tilde {\bm g}}_{i,i} {\tilde {\bm g}}_{i,i}^H\right\},~ \bm G_{i',i}^{(\text R)}= {\mathbb E}\left\{\bm g_{i',i} \bm g_{i',i}^H\right\}, \nonumber\\
&&\bm H_{b,i}^{(\text R)}= {\mathbb E}\left\{\bm h_{b,i} \bm h_{b,i}^H\right\}, ~\forall~ i \in {\cal M}_{\text R},~ i' \in {\cal M}_{\text R}\setminus i.\quad\;
\label{EGH_RUE}
\end{eqnarray}}
\!\!\!From the definition of ${\tilde {\bm g}}_{i,i}$, it is known that ${\tilde {\bm g}}_{i,i}$ consists of i.i.d. Gaussian elements with zero mean and variance given by (\ref{error_RUE}). Hence,
\begin{equation}
\bm E_{i,i}^{(\text R)} = {\text {blkdiag}} \left\{\delta_{k,i} \bm I_N, ~\forall~ k \in {\cal K}_i \right\}.
\label{E_RUE}
\end{equation}
To obtain the explicit expression of $\bm G_{i',i}^{(\text R)}$, denote the set of RRHs serving RUE $i'$ as ${\cal K}_{i'} = \left\{k_1^{i'},\cdots, k_{|{\cal K}_{i'}|}^{i'} \right\}$. Then, according to the definition of $\bm g_{i',i}$, $\bm G_{i',i}^{(\text R)}$ can be rewritten as
\begin{equation}
\bm G_{i',i}^{(\text R)} =
\left[ \begin{array}{ccc}
\!\!\!\!\!\!\left(\bm G_{i',i}^{(\text R)} \right)_{1,1} & \ldots & \left( \bm G_{i',i}^{(\text R)} \right)_{1,|{\cal K}_{i'}|} \\
\vdots & \ddots & \vdots \\
\!\!\!\!\!\!\quad \left( \bm G_{i',i}^{(\text R)} \right)_{|{\cal K}_{i'}|,1} & \;\ldots & \quad \left( \bm G_{i',i}^{(\text R)} \right)_{|{\cal K}_{i'}|,|{\cal K}_{i'}|}
\end{array} \right],
\label{G_RUE}
\end{equation}
where $\left(\bm G_{i',i}^{(\text R)} \right)_{o,z} \triangleq {\mathbb E}\left\{\bm h_{k_o^{i'},i} \bm h_{k_z^{i'},i}^H\right\}, ~\forall~ o,~z \in \left\{1,\cdots,\right.$ $\left. |{\cal K}_{i'}| \right\}$ represents the $o$th row and $z$th column block matrix of $\bm G_{i',i}^{(\text R)}$. From Stage I it is known that for any RUE $i$, only channel vectors $\bm h_{k,i}, \forall k \in {\cal K}_i$ are estimated, while $\bm h_{k,i}, \forall k \in {\cal K}\setminus {\cal K}_i$ are unknown. In addition, different channel vectors are independent with each other. As a result, $\left(\bm G_{i',i}^{(\text R)} \right)_{o,z}$ is given by
\begin{equation}
\left(\!\bm G_{i',i}^{(\text R)} \!\right)_{o,z} \!\!=\! \left\{\!
\begin{array}{ll}
{\hat{\bm h}}_{k_o^{i'},i} {\hat{\bm h}}_{k_o^{i'},i}^H \!\!+\! \delta_{k_o^{i'},i} \bm I_N, & \!\!{\text {if}}~ o=z, ~ k_o^{i'} \in {\cal K}_i, \\
\alpha_{k_o^{i'},i} \bm I_N, & \!\!{\text {if}}~ o=z, ~ k_o^{i'} \not \in {\cal K}_i, \\
0, & \!\!{\text {otherwise}}.
\end{array} \right.
\label{G_RUE1}
\end{equation}
Since for any RUE $i$, $\bm h_{b,i}$ is not estimated, it follows that
\begin{equation}
\bm H_{b,i}^{(\text R)}= \alpha_{b,i} \bm I_B.
\label{H_RUE}
\end{equation}
Substituting (\ref{E_RUE}), (\ref{G_RUE}) and (\ref{H_RUE}) into (\ref{Jensen_RUE}), the lower bound (\ref{rate_lb_RUE}) can be obtained.

Similarly, for BUE $j$, a lower bound on $R_j^{(\text B)}$ can also be derived as (\ref{Jensen_BUE}) shown at the bottom of this page. Denote
\setcounter{equation}{43}
{\setlength\arraycolsep{2pt}%用于缩短等号两边的空格
\begin{eqnarray}
&&\bm E_{b,j}^{(\text B)}= {\mathbb E}\left\{{\tilde {\bm h}}_{b,j} {\tilde {\bm h}}_{b,j}^H\right\},~ \bm G_{i,j}^{(\text B)}= {\mathbb E}\left\{\bm g_{i,j} \bm g_{i,j}^H\right\}, \nonumber\\
&&\bm H_{b,j}^{(\text B)}= {\mathbb E}\left\{\bm h_{b,j} \bm h_{b,j}^H\right\}, ~\forall~ j \in {\cal M}_{\text B},~ i \in {\cal M}_{\text R}.\quad
\label{EGH_BUE}
\end{eqnarray}}
\!\!Since for any BUE $j$, only $\bm h_{b,j}$ is estimated, while channel vectors $\bm g_{i,j}, ~\forall~ i \in {\cal M}_{\text R}$ are unknown, $\bm E_{b,j}^{(\text B)}$, $\bm G_{b,j}^{(\text B)}$ and $\bm H_{b,j}^{(\text B)}$ can thus be readily obtained as follows
{\setlength\arraycolsep{2pt}%用于缩短等号两边的空格
\begin{eqnarray}
\bm E_{b,j}^{(\text B)} &=& \delta_{b,j} \bm I_B,\nonumber\\
\bm G_{i,j}^{(\text B)} &=& {\text {blkdiag}} \left\{\alpha_{k,j} \bm I_N, ~\forall~ k \in {\cal K}_i \right\}, \nonumber\\
\bm H_{b,j}^{(\text B)} &=& {\hat {\bm h}}_{b,j} {\hat {\bm h}}_{b,j}^H + \delta_{b,j} \bm I_B.\quad
\label{EGH_BUE1}
\end{eqnarray}}
\!\!\!\!Substituting (\ref{EGH_BUE1}) into (\ref{Jensen_BUE}), the lower bound (\ref{rate_lb_BUE}) can be obtained.

Since large-scale channel gains and the variance of channel estimation errors, i.e., (\ref{error_RUE}) and (\ref{error_BUE}) are all positive, it can be easily verified that $\bm E_{i,i}^{(\text R)}$, $\bm G_{i',i}^{(\text R)}$, $\bm H_{b,i}^{(\text R)}$, $\bm E_{b,j}^{(\text B)}$, $\bm G_{i,j}^{(\text B)}$ and $\bm H_{b,j}^{(\text B)}$ are all positive definite matrices. Thus, Theorem \ref{theorem1} is proven.

\section{Proof of Theorem \ref{theorem2}}
\label{Appendix_B}
In order to simplify the proof, a special case is first considered. Denote
\begin{equation}
r = \log_2\left(1+ \frac{\left|\bm g^H \bm w\right|^2}{\bm w^H \bm E \bm w + \bm v^H \bm G \bm v + N_0}\right),
\label{rate}
\end{equation}
where $\bm g, \bm w, \bm v \in {\mathbb C}^{C\times 1}$, $\bm E = {\text {diag}} \left\{ \theta_1, \cdots, \theta_C\right\} \in {\mathbb R}_+^{C\times C}$ and $N_0 \in {\mathbb R}_{++}$. $\bm G \in {\mathbb C}^{C\times C}$ is a positive definite Hermitian matrix and admits an eigen-decomposition $\bm G = \bm U \bm \Gamma \bm U^H$, where $\bm U$ is a $C\times C$ dimensional unitary matrix with each column being the eigenvector of $\bm G$ and $\bm \Gamma$ is a diagonal matrix whose diagonal elements are the corresponding eigenvalues of $\bm G$, i.e., $\bm \Gamma_{c,c} = \gamma_c$ \cite{horn2012matrix}. $\theta_c$ and $\gamma_c, ~\forall~ c \in \{1,\cdots, C \}$ are all positive real constants. Let $\bm e_c$ denote a $C$ dimensional vector with one in the $c$th position and zeros elsewhere. $\bm E$ and $\bm \Gamma$ can then be rewritten as $\bm E = \left[ \theta_1 \bm e_1, \cdots, \theta_C \bm e_V\right]$ and $\bm \Gamma = \left[ \gamma_1 \bm e_1, \cdots, \gamma_C \bm e_V\right]$, respectively. Denote ${\bar {\bm v}}= \bm U^H \bm v$. It follows that
\begin{equation}
r = \log_2\left(1+ \frac{\left|\bm g^H \bm w\right|^2}{\sum\limits_{c=1}^C \theta_c \left|\bm e_c^H \bm w\right|^2 + \sum\limits_{c=1}^C \gamma_c \left|\bm e_c^H {\bar {\bm v}}\right|^2 + N_0}\right).
\label{rate1}
\end{equation}

Obviously, (\ref{rate1}) shows that $r$ can be seen as the rate of a mobile user in an equivalent downlink interfering MISO network. The received signal of this user is given by
\begin{equation}
y= \bm g^H \bm w x_0 + \sum\limits_{c=1}^C \sqrt{\theta_c} \bm e_c^H \bm w x_c + \sum\limits_{c=1}^C \sqrt{\gamma_c} \bm e_c^H {\bar {\bm v}} s_c + n,
\label{signal}
\end{equation}
where the first term denotes the desired signal, and the second to the third terms are the interference from other users. $\bm g$, $\sqrt{\theta_c} \bm e_c$ and $\sqrt{\gamma_c} \bm e_c$ denote the channel vectors from all transmitters to the desired receiver. $\bm w$ and $\bm v$ are the transmit beam-vectors. $x_0$, $x_c$ and $s_c$ represent the i.i.d. data symbols with zero mean and unit variance. $n$ is the complex white Gaussian noise with variance $N_0$, i.e., $n \sim {\cal CN}(0,N_0)$. Thus, the MSE of the considered link can be written as
{\setlength\arraycolsep{2pt}%用于缩短等号两边的空格
\begin{eqnarray}
&& {\text {MSE}} = {\mathbb E} \left\{\left|f^H \bm y - x_0 \right|^2 \right\} \nonumber\\
&& = \left|f^H \bm g^H \bm w-1\right|^2 + \sum\limits_{c=1}^C \theta_c \left|f^H \bm e_c^H \bm w\right|^2 \nonumber\\
&&\quad + \sum\limits_{c=1}^C \gamma_c \left|f^H \bm e_c^H {\bar {\bm v}}\right|^2 + N_0 |f|^2 \nonumber\\
&& = \left|f^H \bm g^H \bm w-1\right|^2 + |f|^2 \bm w^H \bm E \bm w+ |f|^2 {\bar {\bm v}}^H \bm \Gamma {\bar {\bm v}} + N_0 |f|^2 \nonumber\\
&& = \left|f^H \bm g^H \bm w-1\right|^2 + |f|^2 \bm w^H \bm E \bm w+ |f|^2 \bm v^H \bm G \bm v + N_0 |f|^2, \nonumber\\
\label{MSE}
\end{eqnarray}}
\!\!and the corresponding MMSE receiver which minimizes (\ref{MSE}) is given by
\begin{equation}
f=\frac{\bm g^H \bm w}{\bm w^H \left(\bm g \bm g^H + \bm E\right) \bm w + \bm v^H \bm G \bm v +N_0}.
\label{MMSE_receiver}
\end{equation}

From (\ref{E_RUE}), (\ref{G_RUE}), (\ref{H_RUE}) and (\ref{EGH_BUE1}), it is known that $\bm E_{i,i}^{(\text R)}$, $\bm H_{b,i}^{(\text R)}$, $\bm E_{b,j}^{(\text B)}$ and $\bm G_{i,j}^{(\text B)}$ are all diagonal matrices as $\bm E$ in (\ref{rate}), and $\bm G_{i',i}^{(\text R)}$ and $\bm H_{b,j}^{(\text B)}$ are both positive definite Hermitian matrices as $\bm G$ in (\ref{rate}). Analogously, Theorem \ref{theorem2} can be readily verified as above.

\bibliographystyle{IEEEtran}
\bibliography{IEEEabrv,MMM}
%\begin{thebibliography}
%\bibitem{BIB17}
%Simulating the SUI Channel Models, IEEE Standard 802.16.3c-01/53, Sep. 2001,http://www.ieee802.org/16/tg3/contrib/802163c-01_53.pdf.
%\end{thebibliography}

% that's all folks
\end{document}